\documentclass[12pt]{article}

\def\hybrid{\topmargin -20pt  \oddsidemargin 0pt
      \headheight 0pt   \headsep 0pt
      \textwidth 6.25in 
      \textheight 9.5in 
      \marginparwidth .875in
      \parskip 5pt plus 1pt   \jot = 1.5ex}

\hybrid

\def\o+{\oplus}

\def\beqa{\begin{eqnarray}}
\def\eeqa{\end{eqnarray}}

\usepackage{amssymb,amsmath,latexsym,axodraw}

\providecommand{\etat}{\tilde{\eta}}

\providecommand{\epsh}{\hat{\epsilon}}
\providecommand{\Gamh}{\hat{\Gamma}}
\providecommand{\eh}{\hat{e}}
\providecommand{\gh}{\hat{g}}
\providecommand{\Rh}{\hat{R}}
\providecommand{\Gh}{\hat{G}}

\providecommand{\ab}{\bar{a}}
\providecommand{\bb}{\bar{b}}
\providecommand{\cb}{\bar{c}}
\providecommand{\db}{\bar{d}}
\providecommand{\eb}{\bar{e}}
\providecommand{\lb}{\bar{l}}
\providecommand{\mb}{\bar{m}}
\providecommand{\nb}{\bar{n}}
\providecommand{\pb}{\bar{p}}
\providecommand{\qb}{\bar{q}}

\providecommand{\ub}{\bar{u}}

\providecommand{\xb}{\bar{x}}

\providecommand{\Ib}{\bar{I}}
\providecommand{\Jb}{\bar{J}}
\providecommand{\Kb}{\bar{K}}

\providecommand{\elfb}{{\bar {11}}}

\providecommand{\nub}{\bar{\nu}}

\begin{document}

\thispagestyle{empty}
\rightline{HU-EP-00/59}
\rightline{hep-th/0012152}
\rightline{\today}
\vspace{2truecm}
\centerline{\bf \LARGE Four-Flux and Warped Heterotic M-Theory}
\centerline{\bf \LARGE Compactifications}

\vspace{1.5truecm}
\centerline{Gottfried Curio $^{a,}$\footnote{curio@physik.hu-berlin.de} 
and Axel Krause $^{b,}$\footnote{krause@physics.uoc.gr}}
\vspace{.6truecm}

{\em
\centerline{$^a$ Humboldt-Universit\"at zu Berlin,}
\centerline{Institut f\"ur Physik, D-10115 Berlin, Germany}}
\vspace{.3truecm}
{\em
\centerline{$^b$ Physics Department, University of Crete,}
\centerline{71003 Heraklion, Greece}}

\vspace{1.0truecm}
\begin{abstract}
In the framework of heterotic M-theory compactified on a Calabi-Yau
threefold 'times' an interval, the relation between geometry and
four-flux is derived {\it beyond first order}.  Besides the case with
general flux which cannot be described by a warped geometry one is
naturally led to consider two special types of four-flux in
detail. One choice shows how the M-theory relation between warped
geometry and flux reproduces the analogous one of the weakly coupled
heterotic string with torsion. The other one leads to a {\it
quadratic} dependence of the Calabi-Yau volume with respect to the
orbifold direction which avoids the problem with negative volume of
the first order approximation.  As in the first order analysis we
still find that Newton's Constant is bounded from below at just the
phenomenologically relevant value. However, the bound does not require
an {\it ad hoc} truncation of the orbifold-size any longer. Finally we
demonstrate explicitly that to leading order in $\kappa^{2/3}$ no
Cosmological Constant is induced in the four-dimensional low-energy
action. This is in accord with what one can expect from supersymmetry.
\end{abstract}

\bigskip \bigskip
\newpage
\pagenumbering{arabic}

\section{Introduction and Summary}

For heterotic M-theory compactified on a $CY_3$ the 4-form
field-strength $G$ does not vanish if higher order corrections in
$\kappa^{2/3}$ are taken into account. The reason is that the boundary
super Yang-Mills (SYM) theories represent magnetic sources which show
up in the Bianchi-identity for $G$ and require a $G$ of order
$\kappa^{2/3}$. Hence, by requiring SUSY which connects via the
Killing-spinor equation metric quantities with the $G$-flux, we
expect warped geometries to arise at this order, which is indeed the
case [\ref{W}]. An interesting interplay arises between the physics of
$G$-fluxes and warped geometry (similar considerations for M- and
F-theory on $CY$ fourfolds can be found in
[\ref{BB}],[\ref{DRS}],[\ref{CPV}],[\ref{GSS}],[\ref{BB2}],[\ref{KB}];
see also [\ref{GSS2}] for dynamical topology changes within heterotic
M-theory). An important phenomenological issue is related to the value
of Newton's Constant $G_N$. From dimensional reduction of heterotic
M-theory on a $CY_3$ with volume $V(x^{11})$ (in the 11-dimensional
metric), one can infer [\ref{W}] (note that these expressions will
become more refined if a non-trivial dependence on internal
coordinates is kept, as will be explored in section 5 below)
\begin{equation}
  G_N = \frac{\kappa^2}{16\pi \langle V\rangle d} \; , \qquad
  \alpha_i = \frac{(4\pi\kappa^2)^{2/3}}{2V_i} \; ,
    \label{NewtonConstant}
\end{equation}
where $\alpha_i$ is the gauge-coupling of the two ($i=1,2$) boundary
$E_8$ SYM theories separated by a distance $d$ and
$V_1=V(0),V_2=V(d)$. Because $G_N$ is related to gravity in the bulk,
we use for its determination a bulk-averaged volume $\langle
V\rangle=\frac{1}{d}
\int_0^d dx^{11} V(x^{11})$. A determination of the warped geometry allows to
calculate $V(x^{11})$ and thereby $\alpha_i$ and $G_N$. This had been
undertaken in [\ref{W}] to linearized order (first order in
$\kappa^{2/3}$) with the result that $V(x^{11})=-ax^{11}+V_0$, where
the slope
$a=\frac{1}{4\sqrt{2}}\int_{CY_3}d^6x\sqrt{g}\,\omega^{lm}\omega^{np}
G_{lmnp}>0$ is controlled by the $G$-flux. Here $\omega_{lm}$ denotes
the K\"ahler-form on $CY_3$. The surprising observation [\ref{W}] has
been that when the linear function $V(x^{11})$ becomes zero, the
corresponding distance for $d$ just gives rise to the correct value
for $G_N$, whereas generically in heterotic string compactifications
$G_N$ is predicted too large by a factor of 400. Placing the second
boundary at that distance means $\alpha_2 \rightarrow \infty$. Hence,
the SYM there becomes strongly coupled and instanton contributions
become relevant, which is the reason why this second boundary
corresponds to a ``hidden'' world rather than our ``observable''
world. The first boundary at $x^{11}=0$ instead allows for a
perturbative SYM on it if $V_1$ is chosen huge enough such that
$\alpha_1 \ll 1$ and consequently can be regarded as our
``observable'' world.

In this context some questions arise
\begin{itemize}
 \item How is the linear behaviour of $V(x^{11})$, which leads to an
 unphysical negative volume beyond a certain distance, changed in the
 full theory, i.e.~beyond the leading $\kappa^{2/3}$ order?  \item
 Does $V(x^{11})$ still keeps its attractive feature of becoming zero
 just at the phenomenological relevant distance? One has to recall
 that actually the linear approximation breaks down at the position of
 the zero. Is Newton's Constant still bounded from below beyond
 leading order?  \item Does the phenomenologically relevant distance
 $d$ get stabilized by an effective potential ?  \item What is the
 relation of the compactification of heterotic M-theory with $G$-flux
 to the weakly coupled heterotic string with torsion [\ref{S}] ?
\end{itemize}
The trouble with the linear behaviour is the following. Though one
expects that quantum corrections will shift the actual value of
$V(x^{11})$ slightly, small distortions of a linear function can never
lift its zero -- they can only shift its $x^{11}$ position slightly,
but the zero remains, as does the problem with the unphysical negative
volume.  Therefore it is important to determine the warp-factors and
thereby $V(x^{11})$ beyond the leading order in $\kappa^{2/3}$, which
we will undertake in this paper.

It may sound surprising how a result beyond order $\kappa^{2/3}$ can
be achieved within the framework of heterotic M-theory whose effective
action is only known to order $\kappa^{2/3}$. Let us therefore briefly
indicate where and in which way features of heterotic M-theory will
enter our analysis.  By imposing supersymmetry, we are going to solve
the gravitino Killing-spinor equation of M-theory. The heterotic
M-theory characteristics enter on the one hand through specific
$G$-fluxes originating from boundary or M5-brane sources and on the
other hand through the chirality condition $\Gamma^{11}\eta=\eta$ on
the susy-variation Majorana-parameter $\eta$. The important point is
that {\it the information which is restricted to order $\kappa^{2/3}$
becomes only relevant if knowledge about the actual source strengths
is required. However, to obtain the functional behaviour of
$V(x^{11})$ this knowledge is not needed}. It suffices to assume that
in the full heterotic M-theory the relevant sources can be localized
in the $x^{11}$ direction (as suggested by the anomaly considerations
of [\ref{HorWitt1}] leading to the two $E_8$ gauge groups), i.e.~they
appear as $dG=\delta(x^{11}-z)S(x^m)\wedge dx^{11}$ in the
Bianchi-identity. Thus, we will be able to answer the first question
posed above. The actual value (and thereby the complete knowledge
about heterotic M-theory beyond $\kappa^{2/3}$ order) for the 4-form
source strength $S$ only becomes indispensable if e.g.~questions about
the precise value of a zero or a minimum of $V(x^{11})$ should be
answered. This is necessary if one wants to quantify the resulting
value for Newton's Constant.

The main results of this paper concerning the questions posed above
are
\begin{itemize}
\item The full dependence of the Calabi-Yau volume on the orbifold
 coordinate is quadratic with a manifest non-negative volume
 throughout the whole interval. The minimum of the parabola is a zero
 which can now be lifted by quantum effects.
\item This zero of the
 nonlinear analysis corrects the zero of the first order analysis by a
 factor two. The critical lower-bound on Newton's Constant in the first order
 approximation is thereby
 further decreased by a factor 2/3. This ameliorates the slight
 discrepancy between the lower bound on Newton's Constant and its
 observed value.
\item The distance $d$ between the walls cannot be stabilized
 by means of an effective potential, which is obtained by integrating
 out the internal dimensions of the eleven-dimensional heterotic
 M-theory action on the warped gravitational background. We
 demonstrate this explicitely at the $\kappa^{2/3}$ level by
 showing the vanishing of the four-dimensional cosmological
 constant.
\item In the case where the volume does not depend on the
 orbifold coordinate one is able to reproduce the relation between
 warp-factor and torsion of the weakly coupled heterotic string.
\end{itemize}

\section{Deriving the Full Relation between Warped Geometry and $G$-Flux in
         Heterotic M-Theory}
Let us consider heterotic M-theory compactified on $CY_3\times
S^1/\mathbb{Z}_2$ with four external coordinates $x^\mu\; ; \,
\mu,\nu,\rho,\hdots=1,2,3,4$
and seven internal coordinates $x^u\; ; \; u,v,w,x,y,z=5, \hdots ,11$. In the
absence of
any $G$-flux (for heterotic M-theory this amounts to considering only the
leading order which is M-theory itself without boundary or M5-brane sources)
the metric solution to the Killing-spinor equation, which describes a
supersymmetry-preserving vacuum, is given by
\begin{equation}
   ds^2=\eta_{\mu\nu}dx^\mu dx^\nu + g_{uv}(x^w)dx^u dx^v \; ,
   \label{Metric}
\end{equation} 
where $g_{uv}$ decomposes into a direct product of the Calabi-Yau metric
$g_{a\bb}$ and the metric $g_{11,11}$ of the
eleventh dimension. Without loss of generality one can set $g_{11,11}=1$.
We will denote the six real Calabi-Yau indices $l,m,n,p,q,\hdots$ while
$\lb,\mb,\nb,\pb,\qb,\hdots$ are the respective flat tangent space indices.
The alternative choice of holomorphic and anti-holomorphic indices will be
denoted $a,b,c\hdots$ and $\ab,\bb,\cb\hdots$.
Genuinely we have to take the boundary sources into account which
require turning on a $G$-flux in the internal directions. This necessitates a
more general metric, for which we choose the warp-factor Ansatz
\begin{alignat}{3}
   ds^2&= \gh_{MN} dx^M dx^N \; ; \; M,N=1,\hdots,11 \notag \\
       &= e^{b(x^n\!\!\!,\,x^{11})}\eta_{\mu\nu}dx^\mu dx^\nu
        + e^{f(x^n\!\!\!,\,x^{11})}g_{lm}(x^n)dx^l dx^m
         +e^{k(x^n\!\!\!,\,x^{11})}dx^{11}dx^{11} \; .
   \label{WarpMetric}
\end{alignat}
It will turn out that the appropriate $G$-flux of the relevant sources can be
accomodated with this Ansatz. The most general Ansatz which allows for
arbitrary $G$-flux compatible with supersymmetry will be considered in the
last section.

The initial Calabi-Yau manifold possesses a closed
K\"{a}hler-form $\omega_{a\bb}$. However, a non-zero $G$-flux entails
a non-trivial internal warp-factor $e^f$, thereby rendering the ``deformed''
K\"{a}hler-form ${\hat\omega}_{a\bb}=e^f\omega_{a\bb}$ non-closed. In this
respect, the warp-factor $e^f$ serves as a measure for the deviation from
K\"{a}hlerness of the internal complex threefold.

To preserve 4-dimensional Poincar\'{e}-invariance, we
set all components of $G$ with at least one external
index to zero\footnote{An external $G_{1234}=\epsilon_{1234}
  \lambda(x^m,x^{11})$ would be compatible with Poincar\'e-symmetry but the
  known sources (boundaries, M5-branes) do not give rise to such a
  flux. Moreover, compatibility with the Bianchi-identity allows only a
  constant $G_{1234}$. Though such a sourceless constant field-strength is
  allowed by the non-compact external spacetime, we will set it to
  zero subsequently.}.
An important point is that in order to preserve supersymmetry
the magnetic sources on the right-hand-side of the Bianchi-identity must be
(2,2,1) forms [\ref{W}]. I.e.~they are forms with two holomorphic,
two anti-holomorphic indices and one $x^{11}$-index. This is clear for the
boundary sources and amounts for the M5-brane sources to an orientation
parallel to the boundaries. Solving the Bianchi-identity, we see
from the fact that the sources are (2,2,1) forms, that only
the components $G_{a\bb c\db}$, $G_{ab\cb 11}$, $G_{a\bb\cb 11}$ can become
non-zero.

\subsection{The Killing-Spinor Equation}
The supersymmetry-variation of the gravitino in low-energy M-theory is given
in the full metric (\ref{WarpMetric}) by
\begin{equation}
  \delta\Psi_I = {\hat D}_I\etat + \frac{\sqrt{2}}{288}\left(\Gamh_{IJKLM}
                 -8\gh_{IJ}\Gamh_{KLM}\right) G^{JKLM}\etat \; , 
  \label{Killing}
\end{equation}
where $\etat=e^{-\psi(x^m,x^{11})}\eta$. Here, $\eta$ is the original
covariantly constant spinor and the exponential-factor accounts for
the correction if $G$-flux is turned on. We will assume $\psi$ to be real
and see later on that
this is indeed compatible with supersymmetry in the warped background.
Subsequently, indices are raised and lowered with the full metric $\gh_{MN}$,
which is also how contractions are performed in (\ref{Killing}).
Setting the variation to zero in order to obtain a supersymmetry preserving
solution, we obtain the Killing-spinor equation.\\[-3mm]

\noindent
{\it Covariant Derivative Contribution}\\[2.5mm]
Let us first deal with the part containing the covariant derivative of
the Majorana-spinor $\etat$. Using the definition of the
spin-connection for the warped-metric\footnote{Here $\Jb,\Kb,\hdots$ denote
  flat 11-dimensional indices.}
\begin{alignat}{3}
  \Omega_{I\Jb\Kb}(\eh)&=\frac{1}{2}
                         \big(\eh_{\Jb}^{\phantom{J}J}
                              {\tilde \Omega}_{IJ\Kb}(\eh)
                             -\eh_{\Kb}^{\phantom{K}K}
                              {\tilde \Omega}_{IK\Jb}(\eh)
                             -\eh_{\Jb}^{\phantom{J}J}
                              \eh_{\Kb}^{\phantom{K}K}
                              \eh^{\Ib}_{\phantom{I}I}
                              {\tilde \Omega}_{JK\Ib}(\eh)
                         \big) \; ,  \\
  {\tilde \Omega}_{IJ\Kb}(\eh)&=  \partial_I \eh_{\Kb J} 
                                - \partial_J \eh_{\Kb I} 
                                      \notag \; ,
\end{alignat}
allows to express the warped spin-connection through the initial one
\begin{alignat}{3}
  \Omega_{\mu\nub\lb}(\eh) &= \frac{1}{2}
           {\eh}_{\lb}^{\phantom{l}m}{\eh}_{\nub\mu}\partial_m b  \; , \qquad
  \Omega_{\mu\nub\elfb}(\eh) = \frac{1}{2}\eh_{\nub\mu}
                 \eh_{\elfb}^{\phantom{11}11}\partial_{11}b       \notag \\
  \Omega_{l\mb\nb}(\eh) &= -{\eh}_{[\mb}^{\phantom{[m}p}
                            {\eh}_{\nb ]l}^{\phantom{[m}}\partial_p f   
                           +\Omega_{l\mb\nb}(e)                   \; , \qquad
  \Omega_{l\mb\elfb}(\eh) = \frac{1}{2}\eh_{\mb l}
                 \eh_{\elfb}^{\phantom{11}11}\partial_{11}f              \\
  \Omega_{11\lb\elfb}(\eh) &= -\frac{1}{2}\eh_{\lb}^{\phantom{l}m}
                 \eh_{\elfb,11}\partial_m k  \notag  \; ,
\end{alignat}
and all other terms are zero. This is done to employ
the covariant constancy $D_I\eta=(\partial_I+\frac{1}{4}
\Omega_{I\Jb\Kb}(e)\Gamma^{\Jb\Kb})\eta=0$ of the initial
spinor-parameter, which brings us to
\begin{alignat}{3}
   dx^I{\hat D}_I\etat = 
             \bigg(&- dx^u \partial_u\psi 
                    + \frac{1}{4}dx^\mu
                      \bigg[ \Gamh_\mu^{\phantom{\mu}l}\partial_l b
                            +\Gamh_\mu^{\phantom{\mu}11}\partial_{11}b
                      \bigg]
                    + \frac{1}{4}dx^l
                      \bigg[
                             \Gamh_l^{\phantom{l}m}\partial_m f
                            +\Gamh_l^{\phantom{U}11}\partial_{11} f
                      \bigg]  \notag \\
                   &+ \frac{1}{4}dx^{11}\Gamh_{11}^{\phantom{11}l}\partial_l k
             \bigg)\etat \; .
\end{alignat}
Let us now specify, that our internal space actually consists of a Calabi-Yau
and a separate eleventh dimension. The positive chirality condition
$\Gamma^{11}\eta=\eta$ on the original space translates into
\begin{equation}
  \Gamh^{11}\etat=e^{-k/2}\etat
  \label{Chirality}
\end{equation}
on the warped space. The condition that we
have a covariantly constant spinor (and its complex conjugate) on the
Calabi-Yau gives $\Gamma^a\eta=0$, $\Gamma_{\ab}\eta=0$ and translates into
$\Gamh^a\etat=0$, $\Gamh_{\ab}\etat=0$. Using these relations plus the
Dirac-algebra $\{\Gamh^a,\Gamh^{\bb}\}=2\gh^{a\bb}$, we end up
with\footnote{It is interesting to look at the Killing-spinor equation with
  general internal $G$-flux but without applying the condition
  of $\mathbb{Z}_2$ invariance, i.e.~$\Gamma^{11}\eta=\eta$, or using the
  Calabi-Yau condition $\Gamma^a\eta = 0$ (which means that
  we are considering an M-theory compactification on a smooth 7-manifold).
  In this case, which will be briefly treated in appendix \ref{SevenMani}, the
  Killing-spinor equation can only be solved by trivial warp-factors and
  a vanishing $G$-flux. Hence, M-theory compactifications on smooth
  7-manifolds neither allow for nontrivial warped metrics nor for nonzero
  internal $G$-flux. This agrees with the result of [\ref{Spence}].} 
\begin{alignat}{3}
   dx^I{\hat D}_I\etat = \bigg\{ 
      &\bigg[-\partial_a(\psi+\frac{f}{4})dx^a
             +\partial_{\ab}(-\psi+\frac{f}{4})dx^{\ab}
             -\partial_{11}\psi dx^{11}\bigg] \notag \\
      +&\bigg[ \frac{1}{4}e^{-k/2}\partial_{11}b\, dx_\mu\bigg]\Gamh^\mu
      +\bigg[ \frac{1}{4}e^{-k/2}\partial_{11}f dx_{\ab}
             -\frac{1}{4}e^{-k/2}\partial_{\ab}k dx_{11}\bigg]\Gamh^{\ab} 
                     \label{CovariantDerivative}     \\
      +&\bigg[ \frac{1}{4}\partial_{\ab}b\, dx_\mu \bigg]\Gamh^{\mu\ab}
      +\bigg[ \frac{1}{4}\partial_{\bb}f dx_{\ab} \bigg]\Gamh^{\ab\bb}
                    \bigg\}\etat                \notag \; .
\end{alignat}
\\[-3mm]

\noindent
{\it $G$-Flux Contributions}\\[2.5mm]
Next, let us deal with the second term in the Killing equation, containing the
$G$-flux. To obtain condense expressions, it proves convenient to parameterize
the three sorts of allowed fluxes by defining
\begin{alignat}{3}
  \alpha &= \omega^{lm}\omega^{np}G_{lmnp}
    \label{Alpha} \\
  \beta_l &= \omega^{mn}G_{lmn11} 
    \label{Beta} \\
  \Theta_{lm} &= G_{lmnp}\omega^{np} 
    \label{GData}  \; .
\end{alignat}
where $\omega_{a\bb}=-ig_{a\bb}$, $\omega^{a\bb}=ig^{a\bb}$ denotes the
K\"ahler-form of the initial Calabi-Yau manifold. The warped metric is related
to the K\"ahler-form by $\gh^{a\bb}=-ie^{-f}\omega^{a\bb}$. Subsequently, we
will make use of
\begin{alignat}{3}
  \gh^{a\bb}\gh^{c\db}G_{a\bb c\db} &= -\frac{1}{4} e^{-2f} \alpha \notag \\
  \gh^{b\cb}G_{l b\cb 11} &=-\frac{i}{2}e^{-f}\beta_l \\
  \gh^{c\db}G_{lm c\db} &=-\frac{i}{2}e^{-f}\Theta_{lm}            \notag
\end{alignat}
to express the occuring contractions through the above defined parameters.
In order to handle the various contractions of $\Gamh$-matrices with the
$G$-flux, it is convenient to evaluate the expressions by first letting the
matrices act on $\etat$ and employ $\gh^{a11}=0$,
$\Gamh^{11}\etat=e^{-k/2}\etat$, $\Gamh^a\etat=0$. Taking $\etat$ as the
ground state, $\Gamh^a$ and $\Gamh^{\ab}$ can be regarded as annihilation and
creation operators, respectively. This leads to some useful identities
(\ref{GammaIdentities}) collected in the appendix. With their help, we
establish the various contractions (\ref{FiveGammaContractions})
of the five-index $\Gamh$-matrices with the $G$-flux and
also the contractions (\ref{ThreeGammaContractions}) of the three-index
$\Gamh$-matrices with $G$. These can be found in the appendix, as well.
Putting all this together, we arrive at the following expression for the
second part of the Killing-spinor equation
\begin{alignat}{3}
  d&x_I\big(\Gamh^{IJKLM}-8\gh^{IJ}\Gamh^{KLM}\big)G_{JKLM}\etat
   = \bigg\{ 3e^{-k/2-f} \left[4i\beta_{\ab}dx^{\ab}+12i\beta_a dx^a-e^{-f}
                              \alpha\, dx_{11}
                        \right] \notag \\ 
            -&3e^{-2f} \alpha\, dx_\mu\Gamh^\mu
            +3e^{-f} \left[-e^{-f}\alpha\, dx_{\ab}+12i
                           \Theta^{\bb}_{\phantom{b}\ab}dx_{\bb}
                           -8i\beta_{\ab}dx^{11}
                     \right]\Gamh^{\ab} 
            -12ie^{-k/2-f} \beta_{\ab}dx_\mu\Gamh^{\mu\ab} \notag \\ 
            +&3e^{-k/2}\left[4ie^{-f}\beta_{\ab}dx_{\bb}
                             -12G^{\cb}_{\phantom{c}\ab\bb 11}dx_{\cb}
                       \right]\Gamh^{\ab\bb}
     \bigg\}\etat   
          \label{GContractions}  \; .
\end{alignat}
\\[-3mm]

\noindent
{\it Complete Killing-Spinor Equation}\\[2.5mm]
Now, the complete Killing-spinor equation can be composed out of the two pieces
({\ref{CovariantDerivative}) and ({\ref{GContractions}) and is given by
\begin{alignat}{3}
  &dx^I{\hat D}_I\etat + \frac{\sqrt{2}}{288}dx^I\left(\Gamh_{IJKLM}
  -8\gh_{IJ}\Gamh_{KLM}\right) G^{JKLM}\etat \notag \\
  =\,&\Bigg( 
      \bigg[\big(-\partial_a\psi-\frac{1}{4}\partial_a f
                 +i\frac{\sqrt{2}}{8}e^{-k/2-f} \beta_a
            \big) dx^a 
           +\big(-\partial_{\ab}\psi+\frac{1}{4}\partial_{\ab}f
                 +i\frac{\sqrt{2}}{24}e^{-k/2-f} \beta_{\ab}
            \big) dx^{\ab} \notag \\
          &-\big( \frac{\sqrt{2}}{96}e^{k/2-2f}\alpha
                 +\partial_{11}\psi
            \big) dx^{11}
       \bigg]
      +\frac{1}{4}\bigg[ e^{-k/2}\partial_{11}b
                        -\frac{\sqrt{2}}{24}e^{-2f}\alpha 
                  \bigg] dx_\mu \Gamh^\mu
      +\frac{1}{4}\bigg[ e^{-k/2}\partial_{11}f dx_{\ab} \notag \\
                       &-\frac{\sqrt{2}}{24}e^{-2f}\alpha dx_{\ab}
      +i\frac{1}{\sqrt{2}}e^{-f}\Theta^{\bb}_{\phantom{b}\ab}dx_{\bb}
      -\big( e^{-k/2}\partial_{\ab}k      
            +i\frac{\sqrt{2}}{3}e^{-f-k}\beta_{\ab}
       \big) dx_{11} 
                  \bigg]\Gamh^{\ab}
      +\frac{1}{4}\bigg[ \partial_{\ab}b \notag \\
                       &-i\frac{\sqrt{2}}{6}e^{-k/2-f}\beta_{\ab}
                  \bigg]dx_\mu \Gamh^{\mu\ab}
      +\frac{1}{4}\bigg[ \partial_{\bb}f dx_{\ab}
                        +i\frac{\sqrt{2}}{6}e^{-k/2-f}\beta_{\ab} dx_{\bb}
                                         \notag \\
     &-\frac{1}{\sqrt{2}}e^{-k/2}G^{\cb}_{\phantom{c}\ab\bb 11}dx_{\cb} 
                  \bigg]\Gamh^{\ab\bb}
    \Bigg)\etat \notag \\
  =\,\,&0 \notag \; .
\end{alignat}

\subsection{The Warp-Factor -- Flux Relations}
Setting the coefficients of the various $\Gamh$-matrices to zero, we have
to distinguish carefully between the imaginary and the real part of the
equations. For arbitrary vectors $A_U,B_U$, the sum $A^aB_a+A^{\ab}B_{\ab}$ is
real, whereas the difference $A^aB_a-A^{\ab}B_{\ab}$ is purely
imaginary. Furthermore $\alpha$ is a real parameter.\\[-5mm] 

\noindent
{\it The} I, {\it $\Gamh^\mu$ and $\Gamh^{\mu\ab}$-Terms}\\[2.5mm]
From the terms proportional to the unit-matrix, $\Gamh^\mu$ and
$\Gamh^{\mu\ab}$ we receive the relations
\begin{alignat}{3}
  8\partial_a \psi &= \partial_a f = -2\partial_a b = i\frac{\sqrt{2}}{3}
                                                   e^{-k/2-f}\beta_a 
              \label{FirstSet} \\
  4\partial_{11} \psi &= -\partial_{11} b = -\frac{\sqrt{2}}{24}
                                                   e^{k/2-2f}\alpha 
              \label{SecondSet} \; .
\end{alignat}
\\[-6mm]

\noindent
{\it The $\Gamh^{\ab}$-Terms}\\[2.5mm]
The terms proportional to $\Gamh^{\ab}$ lead to
\begin{equation}
   \partial_a k = i\frac{\sqrt{2}}{3}e^{-k/2-f}\beta_a \; ,
\end{equation}
which shows that the warp-factors $f$ and $k$ are equal up to an additive
function $F$ depending merely on $x^{11}$
\begin{equation}
  f(x^W,x^{11})=k(x^W,x^{11})+F(x^{11}) \; .
  \label{fhRelation}
\end{equation}
In the following we will set $F(x^{11})$ to zero since it can be eliminated by
a simple reparameterization of $x^{11}$. Furthermore the $\Gamh^{\ab}$ terms
yield the relation
\begin{equation}
  \partial_{11}f = \frac{\sqrt{2}}{24}e^{k/2-2f}\alpha
                  -i\frac{1}{\sqrt{2}}e^{k/2-f}\Theta^{\ab}_{\phantom{a}\ab}
                  \;\; , \;\; \text{no sum over $\ab$}
  \label{ThetaEquation}
\end{equation}
together with
\begin{equation}
  \Theta^{\bb}_{\phantom{b}\ab} = 0 \;\; , \;\; \bb\not=\ab \; .
  \label{Constraint1}
\end{equation}
Note that in (\ref{ThetaEquation}) there is no summation over the
antiholomorphic
indices $\ab$. Hence (\ref{ThetaEquation}) implies the following
isotropy-condition
\begin{equation}
   \Theta^{{\bar 1}}_{\phantom{1}{\bar 1}}= \hdots =
   \Theta^{{\bar n}}_{\phantom{n}{\bar n}} 
   \label{Constraint2} \; ,
\end{equation}
with $n$ the complex dimension of the Calabi-Yau manifold. Using the identity
$\sum_{\ab={\bar 1}}^{{\bar n}}\Theta^{\ab}_{\phantom{a}\ab}=-\frac{i}{2}
e^{-f}\alpha$, it then follows that (\ref{ThetaEquation}) simplifies to
\begin{equation}
  \partial_{11} f = \frac{\sqrt{2}}{4}\left(\frac{1}{6}-\frac{1}{n}\right)
                    e^{k/2-2f}\alpha \; .
   \label{SimplifiedThetaEquation}
\end{equation} 
\\[-6mm]

\noindent
{\it The $\Gamh^{\ab\bb}$-Terms}\\[2.5mm]
Finally the $\Gamh^{\ab\bb}$ terms lead to an equation, which can be
simplified, using the relation for $\partial_a f$ from (\ref{FirstSet}), to
\begin{equation}
  i e^{-f} \beta_{[\ab} dx_{\bb]} = G^{\cb}_{\phantom{c}\ab\bb 11} dx_{\cb}
  \; .
\end{equation}
The component of this equation with $\cb \not= \ab,\bb$ leads to the following
$G$-flux constraint
\begin{equation}
  G^{\cb}_{\phantom{c}\ab\bb 11} = 0 \;\; , \;\; \cb \not= \ab,\bb  \; ,
  \label{Constraint3}
\end{equation}
whereas the $\cb=\ab$ and $\cb=\bb$ components simply reproduce the defining
relation (\ref{Beta}) for $\beta_{\ab}$. 

To summarize, the Killing-spinor equation leads to the set of equations
(\ref{FirstSet}), (\ref{SecondSet}), (\ref{fhRelation}),
(\ref{SimplifiedThetaEquation}) together with the $G$-flux constraints
(\ref{Constraint1}), (\ref{Constraint2}), (\ref{Constraint3}).

We are now in a position to briefly check that our assumption of choosing
$\psi$ real does not lead to inconsistencies. For this purpose it is enough to
show that Im$\,\psi$ is constant, which in particular means that a zero value
can be maintained. Following [\ref{W}], we use the above equation for
$\partial_a \psi$ and obtain
\begin{alignat}{3}
  \gh^{a\bb}\partial_a\partial_{\bb}\text{Im}\psi&=
  \gh^{a\bb}\partial_a\partial_{\bb}\left(\frac{\psi-{\bar\psi}}{2i}\right)
   =
  \frac{\sqrt{2}}{48}\gh^{a\bb}\left[\partial_{\bb}\big(e^{-3f/2}\beta_a\big)
  +\partial_a\big(e^{-3f/2}\beta_{\bb}\big)
                               \right] \notag \\
  &= \frac{\sqrt{2}}{48} e^{-3f/2} D^m\beta_m \notag \; .
\end{alignat}
Employing $D^m \beta_m=0$, which can be obtained from the field equation for
$G$, one establishes that Im$\,\psi$ is a harmonic function on a compact space
and therefore has to be constant.

\section{Implications of the Warped Geometry}
Let us now analyze the above equations in more detail. 
Notice, that up to now we were not forced to specify whether we compactify on
a $CY_2$ or a $CY_3$ -- the complex dimension $n$ of the $CY_n$ entered as a
free parameter. 

Arbitrary $G$-flux parameters $\alpha,\beta$ are only compatible with a pure
warp-factor description of the internal deformed Calabi-Yau in the
6-dimensional case with $n=2$, as we will now see. For $n=2$ we obtain
\begin{equation}
  \partial_{11} f = -\frac{\sqrt{2}}{12}e^{k/2-2f}\alpha \; ,
\end{equation}
which says, together with (\ref{FirstSet}),(\ref{SecondSet}) that
\begin{alignat}{3}
  8\partial_a \psi &= \partial_a f = -2\partial_a b 
              \label{FirstSet2} \\
  8\partial_{11} \psi &= \partial_{11}f = -2\partial_{11} b 
              \label{SecondSet2} 
\end{alignat}
and implies $8\psi=f=k=-2b$. Here the warp-factors depend on both $x^m$
and $x^{11}$. For $n\not= 2$ the $\partial_{11}f$ part receives a different
prefactor and does not allow for this conclusion. Instead -- as we will see
explicitly for the case of $n=3$ below -- one has to set
either (\ref{FirstSet2}) or (\ref{SecondSet2}) to zero to obtain a consistent
solution. 

The difference between the 4- and 6-dimensional cases was also pointed
out in [\ref{W}]. In contrast to $CY_3$ compactifications, a compactification
on $K=K3\times S^1/\mathbb{Z}_2$ to six dimensions allowed the inclusion of a
general $G$-flux without the need to treat the Calabi-Yau coordinates and the
orbifold coordinate differently. This led to a full derivation of the relation
between warp-factors and $G$-flux without the need for a first order
truncation. Starting on $M^{11}={\bf R}^6\times K$ with
\beqa
ds^2=e^b\eta_{\mu\nu}dx^{\mu}dx^{\nu}+e^fg_{uv}dx^u dx^v \; ,
\eeqa
the equation of motion 
\beqa
D^uG_{uvwx}=0
\eeqa 
is solved by the Ansatz (with the $\epsilon^{(0)}$ tensor in the
original metric)
\beqa
G_{uvwx}=-\epsilon^{(0)}_{uvwxy}\partial^y w
\eeqa
together with the condition
\beqa
\triangle^{(0)} w=\mbox{sources}
\eeqa
with the sources derived from the $sources$ in the Bianchi equation
$dG=sources$. The searched for connection between the warp factor and
the $G$-flux then takes the form
\beqa
  e^b&=&(c+2\sqrt{2}w)^{- 1/3}\nonumber\\
  e^f&=&(c+2\sqrt{2}w)^{2/3} \; .
\eeqa
The fact that $e^f=(e^b)^{-2}$ which we also obtain from
(\ref{FirstSet2}),(\ref{SecondSet2}) will again show up in the 4-dimensional
case without dependence on $x^{11}$. The 6-dimensional metric
is related via the decompactification limit to the 11-dimensional extreme
M5-brane metric $ds^2=\Delta^{1/3}\eta_{\mu\nu}dx^{\mu}dx^{\nu}+
\Delta^{-2/3}\delta_{uv}dx^u dx^v$ where $\Delta^{-1}=c+2\sqrt{2}w$
with $w= q/R^3$ and $R=\sqrt{x_ux^u}$ for $G$ taken to be the magnetic field
of a point charge at $x^A=0$ of charge $q$.

But let us now turn to the 4-dimensional case.
If we choose $n=3$, we have
\begin{equation}
  \partial_{11} f = -\frac{\sqrt{2}}{24}e^{k/2-2f}\alpha \; .
\end{equation}
Taking mixed derivatives of $f$ and $b$ this implies that
$\partial_a\partial_{11}f=0$. A non-trivial solution is either obtained from
$\partial_{11}f=0$ or $\partial_a f=0$. The implications of these two cases
will be analyzed in more detail in the following two sections.

\section{Connection between Strong and Weak Coupling}
In this section, we will present the connection to the heterotic string with
torsion [\ref{S}]. 
The choice, $\partial_{11}f=0$, requires $\alpha=0,\beta_a\not= 0$ and
leads to
\begin{equation}
 8\psi(x^m)=f(x^m)=k(x^m)=-2b(x^m)  
\end{equation}
{\it without any dependence on $x^{11}$}. The required sort of fluxes is
obtained by solving the Bianchi-identity $dG=\sum_{i=1}^m \delta(x^{11}-z_i)
S_i(x^m)\wedge dx^{11}$ with $m$ sources by $G=\sum_{i=1}^m \delta(x^{11}-z_i)
P_i(x^m)\wedge dx^{11}$ with $dP_i=S_i$. This type of geometry seems
tailor-made
for a smooth transition to the weakly coupled heterotic string, since any
$x^{11}$ dependence is lost. Indeed, we will now show that the heterotic
M-theory relation between warp-factor and $G$-flux reproduces the
corresponding relation (\ref{WarpTorsion}) for the heterotic string with
torsion. 

The warp-factor belonging to the 4-dimensional external part multiplies the
Minkowski-metric -- both in the string and the M-theory case -- and is
therefore fixed in the sense that one does not have to take into account
further coordinate-reparameterizations for a comparison. Let us therefore
start with the relation between external warp-factor and $G$-flux by using
(\ref{FirstSet}) for $\partial_a b$ plus $f=k=-2b$ and employing the
definition of $\beta_a$ to obtain
\begin{equation}
  \partial_a (e^{-b}) = -\frac{\sqrt{2}}{3}{{G_{ab}}^b}_{11} \; .
\end{equation}
The contraction on the right-hand-side is with respect to $\gh^{b\cb}$.
To compare M-theory with string-theory [\ref{WittenM}] one has to perform an
overall Weyl-rescaling involving the dilaton,
$g^\sigma_{MN}=e^{2\phi/3}\gh_{MN}$, which brings us to the string-frame. Here
we have
\begin{alignat*}{3}
  \partial_a (e^{-b}) &= -\frac{\sqrt{2}}{3}e^{\frac{2\phi}{3}}
                          {{G_{ab}}^b}_{11} \; ,
                                                                      \\
  ds^2 &= e^{b+\frac{2\phi}{3}}\eta_{\mu\nu}dx^\mu dx^\nu + \hdots \; .
\end{alignat*}
Finally, let us go over to the 10-dimensional Einstein-frame via
$g^E_{AB}=e^{-\phi/2}g^\sigma_{AB}$ in which we obtained the heterotic string
relation between warp-factor and torsion. We thus arrive at
\begin{alignat}{3}
  \partial_a (e^{-b}) &= -\frac{\sqrt{2}}{3}e^{\frac{\phi}{6}}
                          {{G_{ab}}^b}_{11} \; ,
       \label{WarpFluxM}                                             \\
  ds^2 &= e^{b+\frac{\phi}{6}}\eta_{\mu\nu}dx^\mu dx^\nu + \hdots \; ,
\end{alignat}
where again the contraction is performed with the metric of the actual frame,
$(g^E)^{b\cb}$. A comparison of the above metric with the heterotic string
metric (\ref{StromingerMetric}) shows that we have to identify $2\phi$ with
$b+\phi/6$, which gives
\begin{equation}
  b=\frac{11}{6}\phi \; .
\end{equation}
If we use this in (\ref{WarpFluxM}), we receive the heterotic M-theory
warp-factor -- flux relation
\begin{equation}
  \partial_a (e^{-2\phi}) = -\frac{4\sqrt{2}}{11} {{G_{ab}}^b}_{11} \; . 
\end{equation}
Setting ${{G_{ab}}^b}_{11}$ equal to ${H_{ab}}^b$ up to some constant
normalization factor, we see that indeed the relation between warp-factor and
torsion of the heterotic string (see appendix \ref{StringCase} for
relevant facts about the heterotic string with torsion and a derivation of the
following formula in that context)
\begin{equation}
  \partial_a(e^{-2\phi})=-\frac{1}{2}{H_{ab}}^b \; .
\end{equation}
can be reproduced from heterotic M-theory including $G$-flux. This represents a
non-trivial check on the duality between the strongly and the weakly coupled
heterotic regions in the presence of torsion.

The choice of fluxes treated in this subsection leads to a Calabi-Yau volume
which does not depend on $x^{11}$. Moreover due to the deformation with the
warp-factor $e^f$ the K\"ahler-form is
no longer closed (cf.~eq.~(\ref{nonKaehler})).
In addition the Ricci-tensor for the internal six-manifold becomes
\begin{equation}
  R_{a\bb}(e^f g_{mn}) = R_{a\bb}(g_{mn}) + g_{a\bb}g^{c\db}
                          \left( 2\partial_c f \partial_{\db} f
                                + \partial_c \partial_{\db} f
                          \right)
                        -\partial_a f \partial_{\bb} f
                        +2\partial_a \partial_{\bb} f \; , 
     \label{InternalRicci}
\end{equation}
where the derivatives of $f$ are determined through the $G$-flux by
(\ref{FirstSet}). Though $R_{a\bb} (g_{mn})=0$ due to the Ricci-flatness of
the initial Calabi-Yau space, we recognize that in the presence of $G$-flux
the internal six-manifold also looses its property of being Ricci-flat.

\section{The Analysis of the Warped Geometry in the Strong Coupling Case
         and Newton's Constant}
The second choice, $\partial_a f=0$, requires $\alpha\not= 0, \beta_a=0$ and
implies
\begin{equation}
   4\psi(x^{11})=f(x^{11})=k(x^{11})=-b(x^{11})
\end{equation}
{\it without any $x^m$ dependence}. The necessary non-vanishing
$G_{a\bb c\db}$ and vanishing $G_{ab\cb 11}$ are obtained by solving the
Bianchi-identity $dG=\sum_{i=1}^m \delta(x^{11}-z_i) S_i(x^m)\wedge dx^{11}$
through\footnote{The Heaviside step-function $\Theta(x)$ is defined by
  $\Theta(x<0)=0$ and $\Theta(x>0)=1$.}
$G=\sum_{i=1}^m \Theta(x^{11}-z_i) S_i(x^m)$. Again $S_i(x^m)$ is a
closed 4-form representing the strength of the $i^{th}$ magnetic source.

Note that for $\alpha\not= 0,\beta_a=0$ the internal
six-manifold remains K\"ahler. This is due to the fact that the warp-factor
$f$ does not depend on $x^m$. Furthermore, we see from the general formula
(\ref{InternalRicci}) for the Ricci-tensor that in this case the six-manifold
also keeps its property of being Ricci-flat. In other words the six-manifold
is still a Calabi-Yau space with volume depending on the ``parameter''
$x^{11}$.

\subsection{The volume dependence on the orbifold direction}
The volume of the Calabi-Yau, as measured by the warped metric, is given by
$V(x^{11})=\int d^6x\sqrt{\gh_{CY_3}}=e^{3f}\int d^6x\sqrt{g_{CY_3}}$.
The decisive part, which is responsible for the variation of the volume with
$x^{11}$, is the factor $e^{3f}$. For its determination, we use $f=k$ and the
equation for $\partial_{11}f$
\begin{equation}
  \partial_{11}(e^{3f/2}) = -\frac{1}{8\sqrt{2}}\,\alpha
  \label{fWarp}
\end{equation}
which is solved by
\begin{equation}
  e^{3f(x^{11})/2} = e^{3f(0)/2}
                     -\frac{1}{8\sqrt{2}}\int_0^{x^{11}}dz\,\alpha(z)
                     \; .
\end{equation}
Notice that $\alpha$ does not depend on the Calabi-Yau coordinates, which can
be easily seen by acting with $\partial_a$ on (\ref{fWarp}) and taking into
account that $\partial_a f=0$. Hence the variation of the Calabi-Yau volume
with $x^{11}$ is given by
\begin{equation}
  V(x^{11})=\Big( 1
            -\frac{1}{2\sqrt{2}}\,\omega^{a\bb}\omega^{c\db}\int_0^{x^{11}}
             dz\,G_{a\bb c\db}(x^m,z)
            \Big)^2 
            V_1\; ,
\end{equation}
where $V_1=\int d^6x\sqrt{g_{CY_3}}$ is the Calabi-Yau volume in the
initial metric. The integration constant $e^{3f(0)/2}$ has been set to 1 to
obtain a smooth transition from $V(x^{11})$ to $V_1$ in case that we turn off
any $G$-flux. The only assumption about the full heterotic M-theory that we
will have to make is that the sources can still be localized at $x^{11}=z_i$
in the eleventh direction, i.e.~that the Bianchi-identity possesses the form
$dG=\sum_{i=1}^m \delta(x^{11}-z_i) S_i(x^m) \wedge dx^{11}$. Its solution
$G=\sum_{i=1}^m \Theta(x^{11}-z_i) S_i(x^m)$ then leads to the following
behaviour of the Calabi-Yau volume
\begin{equation}
  V(x^{11}) = \Big( 1-\sum_{i=1}^m
                    (x^{11}-z_i) \Theta(x^{11}-z_i) {\cal S}_i
              \Big)^2 V_1 \; ,
\end{equation}
where ${\cal S}_i=\frac{1}{2\sqrt{2}}\omega^{a\bb}\omega^{c\db}
(S_i)_{a\bb c\db} (x^m)$. Thus we get the remarkably simple result that
{\it in the full treatment} the linear behaviour of the first order
approximation gets replaced by {\it a quadratic behaviour}.

For the simplest case with only the two boundary sources at $z_1=0,z_2=d$, we
obtain $\alpha=8\sqrt{2}\Theta(x^{11}){\cal S}_1$ with ${\cal S}_1$
representing the magnetic source of the ``visible'' boundary. This gives the
warp-factor (beyond $x^{11}=x_0^{11}=1/{\cal S}_1$, where the right-hand-side
(rhs) becomes negative, we assume an analytic continuation of the
left-hand-side through the rhs)
\begin{equation}
  e^{3f(x^{11})/2} = 1-{\cal S}_1 x^{11} \; ,
  \label{AnalyticCont}
\end{equation}
which leads to the following volume dependence (see fig.\ref{VolumeVar})
\begin{equation}
  V(x^{11}) = \left( 1-{\cal S}_1 x^{11} \right)^2 V_1 \; .
\end{equation} 
            \setcounter{figure}{0}
            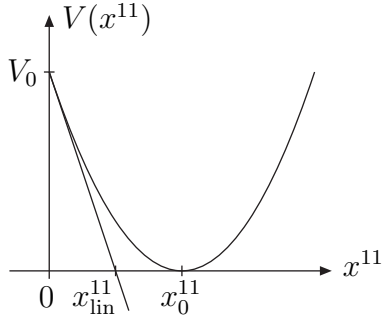
\begin{figure}[t]
              \begin{center}
               \begin{picture}(100,100)(0,-5)
                  \LongArrow(5,2)(5,100)

                  \Line(5,80)(35,-10)
                  \Curve{(5,80)(30,23.75)(55,5)(80,23.75)(105,80)}

                  \LongArrow(-10,5)(110,5)
                  \Line(55,3)(55,7)
                  \Line(3,80)(7,80)                
                  \Line(30,3)(30,7)

                  \Text(-5,80)[]{$V_0$}
                  \Text(4,-5)[]{$0$}
                  \Text(22,-5)[]{$x^{11}_{\text{lin}}$}
                  \Text(124,8)[]{$x^{11}$}
                  \Text(55,-5)[]{$x_0^{11}$}
                  \Text(27,100)[]{$V(x^{11})$}
               \end{picture}
               \caption{The quadratic dependence of the Calabi-Yau volume on
                        the orbifold direction in the full geometry and its
                        linear approximation to order $\kappa^{2/3}$. If
                        higher order contributions are negligible then the
                        linear approximation is valid for small $x^{11}$. The
                        left boundary corresponds to the ``visible'' world.} 
               \label{VolumeVar}
              \end{center}
            \end{figure}
Moreover -- as becomes clear from the figure --
with the quadratic volume behaviour {\it tiny quantum effects are
now able to resolve the zero volume} as opposed to the linearized case
(cf.~in this respect also [\ref{BG}],[\ref{StromLoop}]). We believe
that the perfect square structure of the volume and thereby its zero
at the minimum is related to supersymmetry. The reason why $e^{3f/2}$
is linear in $x^{11}$ can be traced back through the derivation of the
warp-factor--flux relations to the fact, that we had imposed the
condition (\ref{Chirality}). But this condition is
nothing but the chirality condition $\Gamma^{11}\eta = \eta$ for the
initial fermionic parameter $\eta$, which had to be imposed to preserve $N=1$
supersymmetry in ten dimensions.
On a more qualitative level one may argue that in general a solution of
the Einstein equations is a solution to a second order differential equation
with two integration constants. One gets a subclass of supersymmetry-preserving
solutions by solving the Killing-Spinor equation instead. This special subclass
exhibits in contrast only one integration constant. We suppose that the
additional integration constant in the non-susy case parameterizes the
ordinate position of the locus of minimal volume and becomes zero in the
supersymmetric limit. This indicates that the zero could be lifted, not only
by quantum corrections, but also classically by an appropriate Susy-breaking.

Here and in the following the right boundary will not be depicted -- it would
cut off the solution at some finite distance $d$.
To determine the actual value of ${\cal S}_1$
(and thereby Newton's Constant, see below) would require
the actual knowledge of heterotic M-theory to all orders in
$\kappa^{2/3}$. As we see, it is possible to parameterize our
ignorance about these higher-order terms by ${\cal S}_1$. 
In the phenomenological relevant case where ${\cal S}_1>0$ (a positive
${\cal S}_1$
is in accordance with the leading order approximation), a zero volume develops
at $x_0^{11}=1/{\cal S}_1$.

The $x^{11}$ dependence of the full metric reads\footnote{It is
interesting to compare this solution to the domain-wall solution which
arises in the effective 5-dimensional M-theory [\ref{BG}]. In the
11-dimensional decompactification limit it is given by (use eq.~(4.19)
of [\ref{BG}] and set $H_1=H_2=H_3=H(y)$)
\begin{equation}
  ds^2 =  \frac{1}{H} \eta_{\mu\nu} dx^\mu dx^\nu
        + H (d\omega_1+d\omega_2+d\omega_3 + Hdy^2) \; .
    \label{BGSolution}
\end{equation}
Here $d\omega_i$ are 2-dimensional line-elements spanning the internal
six dimensions. Comparing both external 4-dimensional parts gives $H
\sim V^{1/3}$, while from a comparison of the eleventh coordinate
parts we gain the reparameterization $\sqrt{H}dy = dx^{11}$. Noting
that $H\sim y$ [\ref{BG}], we obtain $x^{11}\sim y^{3/2}$ and thus
$H\sim (x^{11})^{2/3}$. Therefore (\ref{WarpSolution}) and
(\ref{BGSolution}) show the same $x^{11}$ dependence for both the
external and the internal parts. The coordinate $x^{11}$ which we are
using instead of say $y$ is distinguished by simple Bianchi-identities
of the form $dG =
\delta(x^{11}-z_i)S(x^m)\wedge dx^{11}$.}
\begin{equation}
  ds^2 =  \left(1-{\cal S}_1 x^{11}\right)^{-2/3} \eta_{\mu\nu}dx^\mu dx^\nu
        +\left(1-{\cal S}_1 x^{11}\right)^{2/3} 
         \left( g_{lm}(x^n)dx^l dx^m + dx^{11}dx^{11} \right)
    \label{WarpSolution}
\end{equation}
or in terms of the Calabi-Yau volume
\begin{equation*}
  ds^2 =  \left(\frac{V(x^{11})}{V_1}\right)^{-1/3} \eta_{\mu\nu}dx^\mu dx^\nu
        + \left(\frac{V(x^{11})}{V_1}\right)^{1/3} 
          \left( g_{lm}(x^n)dx^l dx^m + dx^{11}dx^{11} \right) \; .
\end{equation*}

\subsection{Extracting the First-Order Results}
Finally, one would like to see the transition of the full expression for
$V(x^{11})$ to the linearized expression which was derived in [\ref{W}]. For
this purpose one has to expand the
sources into a power-expansion in $\kappa^{2/3}$. If we are only interested in
sources coming from the boundary, then we know that they start at the order
$\kappa^{2/3}$ 
\begin{equation}
   {\cal S}_1 = {\cal S}_1^{(1)}\kappa^{2/3}
               +{\cal S}_1^{(2)}\kappa^{4/3}
               +\hdots \; .
\end{equation}
For the first order approximation, we have to truncate this series after the
first term, which indeed gives rise to a linear volume dependence
\begin{equation}
   V(x^{11})=\left(1-2{\cal S}_1^{(1)}\kappa^{2/3}x^{11}\right) V_1
             +{\cal O}(\kappa^{4/3})
    \label{LinearApproximation}
\end{equation}
as found in [\ref{W}]. If we now read off the zero of $V(x^{11})$, we get
$x^{11}_{\text{lin}}=1/(2{\cal S}_1^{(1)}\kappa^{2/3})$ in the linearized
case, while the full solution gives a different first order zero
\begin{equation}
  x^{11}_0=1/({\cal S}_1^{(1)}\kappa^{2/3}) + {\cal O}(\kappa^{4/3}) \; .
\end{equation}
This little puzzle is resolved by noticing that the linear approximation
(\ref{LinearApproximation}) holds true only as long as
${\cal S}_1^{(1)}\kappa^{2/3}x^{11}\ll 1$ (plus similar conditions for the
higher ${\cal S}_1^{(i)},\; i\ge 2$ contributions). Because at the
position of the zero, we face ${\cal S}_1^{(1)}\kappa^{2/3}x^{11}_{\text{lin}}
\approx {\cal S}_1^{(1)}\kappa^{2/3}x^{11}_0=1$, the linear approximation
(\ref{LinearApproximation}) breaks down and cannot be used to
determine reliably the zero of $V(x^{11})$. Therefore, in contrast to
the first order analysis, the actual zero at the first order level
becomes larger by a factor 2
\begin{equation}
  x^{11}_0 = 2 x^{11}_{\text{lin}} \; .
\end{equation}
This shows that if we place the ``hidden'' boundary not very close
(such that ${\cal S}_1^{(1)}\kappa^{2/3}d$ is not much smaller
than 1) to the ``visible'' boundary, we are forced to take into account
higher order terms in $\kappa^{2/3}$. In particular this applies for
the phenomenologically interesting region around the volume zero.
\begin{figure}[t] \begin{center} \begin{picture}(400,120)(0,-5)
\Text(10,120)[]{a)} \LongArrow(35,2)(35,100)

                  \Curve{(35,80)(60,23.75)(85,5)(110,23.75)}
                  \Curve{(110,23.75)(116,60)(118,80)}

                  \LongArrow(20,5)(140,5)
                  \Line(85,3)(85,7)
                  \Line(33,80)(37,80)                
                  \Line(110,3)(110,7)

                  \Text(25,80)[]{$V_0$}
                  \Text(34,-5)[]{$0$}
                  \Text(154,8)[]{$x^{11}$}
                  \Text(85,-5)[]{$x_0^{11}$}
                  \Text(57,100)[]{$V(x^{11})$}
                  \Text(110,-6)[]{$z_{M5}$}

                  \Text(255,120)[]{b)}
                  \LongArrow(280,2)(280,100)

                  \Curve{(280,80)(290,47)(305,23.75)}
                  \Curve{(305,23.75)(320,5)(335,23.75)(345,57.08)(350,80)}

                  \LongArrow(255,5)(355,5)
                  \Line(278,80)(282,80)                
                  \Line(320,3)(320,7)
                  \Line(305,3)(305,7)

                  \Text(270,80)[]{$V_0$}
                  \Text(280,-5)[]{$0$}
                  \Text(369,8)[]{$x^{11}$}
                  \Text(302,100)[]{$V(x^{11})$}
                  \Text(305,-6)[]{$z_{M5}$}
                  \Text(325,-5)[]{${\tilde x}_0^{11}$}
               \end{picture}
               \caption{The figure shows the volume dependence in the
                        presence of an additional M5-brane at $z_{M5}$, where
                        we assume a positive M5-brane source ${\cal S}_{M5}$.
                        In a) 
                        the situation for $x_0^{11}<{\tilde x}_0^{11}<z_{M5}$
                        is depicted while b) shows the behaviour for
                        $x_0^{11}>{\tilde x}_0^{11}>z_{M5}$. The second
                        boundary at $x^{11}=d$ is not depicted -- it would
                        truncate the solution at the finite distance $d$.} 
               \label{VolumeVarM5}
              \end{center}
            \end{figure}
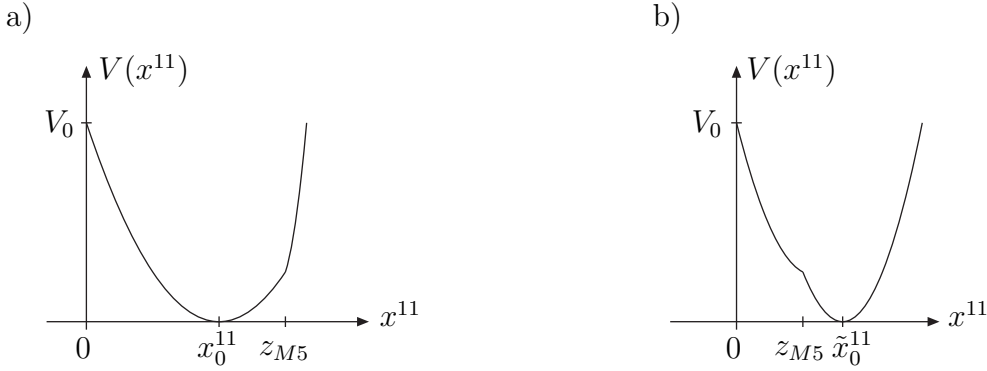

\subsection{Inclusion of M5-branes}
Let us briefly consider the case with three sources -- two from the
boundaries ${\cal S}_1, {\cal S}_2$ plus a further one ${\cal S}_{M5}$ from an
M5-brane placed in between at $z_{M5}$. With
$\alpha=8\sqrt{2} [ \Theta(x^{11}){\cal S}_1 + \Theta(x^{11}-z_{M5})
{\cal S}_{M5} ]$ we get a warp-factor
\begin{equation}
  e^{3f(x^{11})/2} = 1-x^{11}{\cal S}_1
                     -(x^{11}-z_{M5})\Theta(x^{11}-z_{M5}){\cal S}_{M5}
\end{equation}
and the following volume dependence (see fig.\ref{VolumeVarM5})
\begin{equation}
 V(x^{11}) = \left\{ \begin{array}{ccc} \left(1-{\cal S}_1
 x^{11}\right)^2\, V_0 & ; & x^{11} < z_{M5} \\ \big(1-({\cal
 S}_1+{\cal S}_{M5})x^{11}+{\cal S}_{M5} z_{M5}\big)^2\, V_0 & ; &
 x^{11}\ge z_{M5} \end{array} \right.
\end{equation}
The zero of the parabola for $x^{11}\ge z_{M5}$ lies at ${\tilde x}_0^{11}
=(1+{\cal S}_{M5} z_{M5})/({\cal S}_1+{\cal S}_{M5})$.
Thus we see that an additional M5-brane will increase or decrease the slope of
the volume parabola depending on whether it contributes a positive or negative
${\cal S}_{M5}$.

\subsection{Newton's Constant}
Let us briefly recall the evaluation of Newton's Constant in the first
order analysis [\ref{W}]. It is obtained by a dimensional reduction
procedure together with an average Calabi-Yau volume $\langle V
\rangle = \frac{1}{d}\int_0^d dx^{11} V(x^{11})$ through
(\ref{NewtonConstant}). It had been found [\ref{W}] that
\begin{equation}
 G_N \ge G_N^{crit,lin} \equiv \frac{\kappa^2}{8\pi V_1 x_{\text{lin}}^{11}}
      = \frac{\alpha_1^2}{16\pi^2}
        \left| \int_{CY_3} \omega \wedge 
               \left( \frac{\text{tr} F^2-\frac{1}{2}\text{tr} R^2}{8\pi^2}
               \right)
        \right|       \; .
\end{equation}
An approximation of the integral by $1/M_{\text{GUT}}^2$ delivers (with
$M_{\text{GUT}}=2\times 10^{16}\text{GeV and } \alpha_1 = 1/25$) the
lower bound
\begin{equation}
  G_N^{crit,lin} \simeq \frac{1}{(6.3 \times 10^{18}\text{GeV})^2} \; ,
\end{equation}
which is only slightly bigger than the actual phenomenological value
$G_N^{phen}=1/(1.3\times 10^{19}\text{GeV})^2$.

In order to obtain effective 4-dimensional entities like $G_N$ or the
``hidden'' gauge-coupling $\alpha_2$ beyond the linear approximation,
we have to keep the full $x^{11}$ dependence of the fields and
integrate out the seven compact dimensions (in contrast to performing
a simple dimensional reduction).  Starting with the 11-dimensional
Einstein-Hilbert term
\begin{equation}
  S = -\frac{1}{2\kappa^2}\int d^4x \int d^6x \int dx^{11}
       \sqrt{\gh^{(4)}}\sqrt{\gh_{CY_3}}\sqrt{\gh_{11}}
       R(x^{11})
\end{equation}
we can use the full metric information (\ref{WarpSolution}) to
explicitly integrate over $x^{11}$ and the Calabi-Yau coordinates (the
integration over the Calabi-Yau coordinates will be contained in the
volume $V(x^{11})$). To gain a non-vanishing 4-dimensional curvature
scalar, we consider slight perturbations around the flat 4-dimensional
spacetime, $\eta_{\mu\nu}\rightarrow g^{(4)}_{\mu\nu}$. Keeping only
the 4-dimensional curvature-scalar part from the reduction of the
11-dimensional curvature scalar, $R(\gh_{MN})=
e^{-b}R(g^{(4)}_{\mu\nu}) + \hdots$, we match the effective
4-dimensional Einstein-Hilbert action
\begin{equation}
  S^{(4)} = -\frac{1}{16\pi G_N}\int d^4 x \sqrt{g^{(4)}}
  R(g^{(4)}_{\mu\nu})
\end{equation}
with the following effective Newton's Constant (in the ``downstairs''
picture [\ref{HorWitt2}] , which we are employing, an additional
factor of two multiplying the integral over $x^{11}$ has to be taken
into account)
\begin{equation}
   G_N = \frac{\kappa^2}{16\pi \int_0^d dx^{11} e^{b(x^{11})/2}
   V(x^{11})}  \; .
\end{equation}
Note that it is not only the volume of the internal 7-fold but also an
additional warp-factor stemming from the external 4-dimensional metric
which enters the expression for the effective $G_N$.  Though the
CY-volume is manifestly positive, the square-root of the warp-factor
$e^b$ becomes negative beyond $x^{11}_0$ if we use the analytic
continuation mentioned before (\ref{AnalyticCont}). Notice, however,
that it is only the square root of the warp-factors which is continued
into the negative region. The warp-factors themselves stay positive
under this continuation, as can be seen explicitly by comparing
(\ref{WarpMetric}) with (\ref{WarpSolution}). Also physical entities
like the Ricci-tensor or the Riemann curvature scalar are well-behaved
under this continuation.

If we now insert the known warp-factor and volume dependences, we
arrive at
\begin{equation}
  G_N = \frac{2}{3} G_N^{crit} 
        \frac{1}{\left[ 1 - \left(1-\frac{d}{x_0^{11}}\right)^{8/3}
        \right]}
      \ge \frac{2}{3} G_N^{crit} \; ,
\end{equation}
where $G_N^{crit}=\frac{\kappa^2}{4\pi V_1 x_0^{11}}$ approaches the
lower bound of the first order analysis $G_N^{crit,lin}$ if $x^{11}_0
\rightarrow 2 x^{11}_{\text{lin}}$. Notice that $\frac{2}{3}
G_N^{crit}$ places a lower bound on Newton's Constant which depends
via $x^{11}_0=1/{\cal S}_1$ on the source strength. In the first order
approximation this lowers the previously [\ref{W}] obtained lower
bound by a further factor $2/3$, which is welcomed for
phenomenological purposes.

The dependence of $G_N$ on $d$ is symmetric around
the zero-position $x_0^{11}$ (see fig.\ref{GNewtonAndVol7}a). 
            \begin{figure}[t]
              \begin{center}
               \begin{picture}(400,120)(0,-5)
                  \Text(10,120)[]{a)}
                  \LongArrow(35,2)(35,105)
                  \LongArrow(20,5)(140,5)
                  \Text(34,-5)[]{$0$}
                  \Text(151,8)[]{$d$}
                  \Text(50,105)[]{$G_N$}

                  \Curve{(36,96.6)(38,46.6)(40,36.7)(45,29.35)(50,27)
                         (55,25.95)(65,25.15)(75,25)}
                  \Curve{(75,25)(85,25.15)(95,25.95)(100,27)(105,29.35)
                         (110,36.7)(112,46.6)(114,96.6)}

                  \DashLine(35,25)(115,25){3}
                  \Text(15,25)[]{$\frac{2}{3}G_N^{crit}$}
                  \Line(75,3)(75,7)
                  \Text(75,-5)[]{$x_0^{11}$}
                  \Line(115,3)(115,7)
                  \Text(115,-5)[]{$2 x_0^{11}$}

                  \Text(255,120)[]{b)}
                  \LongArrow(280,2)(280,105)
                  \LongArrow(265,5)(385,5)
                  \Text(279,-5)[]{$0$}
                  \Text(396,8)[]{$d$}
                  \Text(295,105)[]{$V_7$}
                  \DashLine(280,80)(360,80){3}
                  \Text(260,80)[]{$\frac{3}{5}V_1 x_0^{11}$}

                  \Curve{(280,5)(281,11.08)(285,32)(290,51.5)(300,72.5)
                         (310,79.25)(320,80)}
                  \Curve{(320,80)(330,79.25)(340,72.5)(350,51.5)(355,32)
                         (359,11.08)(360,5)}

                  \Line(320,3)(320,7)
                  \Text(320,-5)[]{$x_0^{11}$}
                  \Line(360,3)(360,7)
                  \Text(360,-5)[]{$2 x_0^{11}$}
               \end{picture}
               \caption{The dependence of $G_N$ on $d$ is shown in
                        figure a). Qualitatively (modulo an external
                        warp-factor) this can be understood from the
                        corresponding variation of the seven-fold $CY_3\times
                        S^1/\mathbb{Z}_2$ volume
                        $V(\text{7-fold})$ with $d$, figure b). Note that the
                        decrease of $V(\text{7-fold})$ beyond $x_0^{11}$
                        results from the analytic continuation of the
                        warp-factors into the negative region.}
               \label{GNewtonAndVol7}
              \end{center}
            \end{figure}
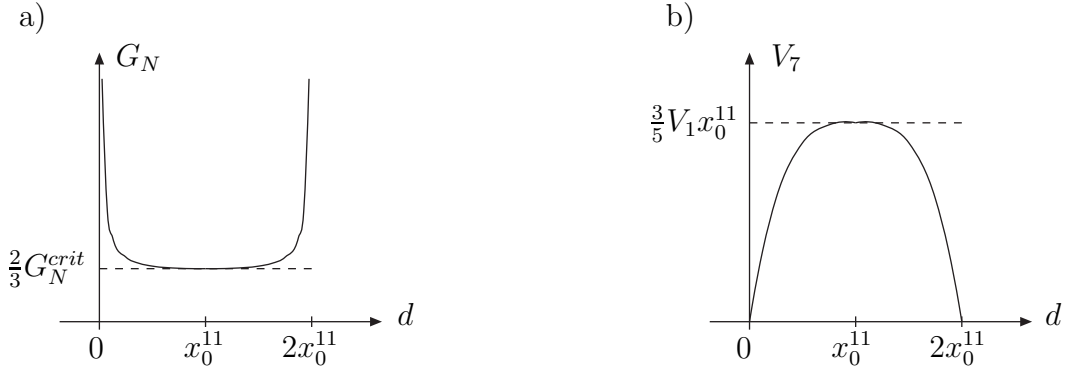
This implies that the effective Newton's Constant does not only diverge  
at $d=0$ (as it already does at the zeroth-order approximation
(\ref{NewtonConstant})) but also at $d=2x_0^{11}$. Hence, the length
of the orbifold-interval has to be upper-bounded by $d\le 2x_0^{11}$.
Notice furthermore, that qualitatively the inverse of the
seven-fold $CY_3\times S^1/\mathbb{Z}_2$ volume
\begin{alignat}{3}
     V(\text{7-fold})
   &= \int d^6 x \,2\int_0^d dx^{11}\sqrt{\gh_{CY_3}} \sqrt{\gh_{11,11}} 
    = 2\int_0^d dx^{11} e^{k(x^{11})/2} V(x^{11})    \notag \\
   &= \frac{3}{5} V_1 x_0^{11}
     \left[ 1 - \left(1-\frac{d}{x_0^{11}}\right)^{10/3} \right]
\end{alignat}
reflects the main features of $G_N$ (see fig.\ref{GNewtonAndVol7}b),
as one would expect from the zeroth-order formula
(\ref{NewtonConstant}). However, quantitatively they differ by the
contribution of an additional external warp-factor $e^b$ which appears
under the $x^{11}$ integral of $G_N$.

Both the Calabi-Yau volume and the warp-factors are symmetric with
respect to $x_0^{11}$. Therefore it is clear that this property also holds for
the gauge-coupling (as a function of $d$) of the hidden boundary. Let
us briefly derive the corresponding expression for the hidden
gauge-coupling. Starting with the 11-dimensional gauge-kinetic term
\begin{equation}
  S = -\frac{1}{8\pi(4\pi\kappa^2)^{2/3}} \int
  d^{10}x\sqrt{\gh^{(4)}}\sqrt{\gh_{CY_3}} \gh^{AC}\gh^{BD} F^a_{AB}
  F^a_{CD}
\end{equation} 
we focus on the 4-dimensional $F^a_{\mu\nu}$ components, which are
supposed to depend solely on $x^\mu$. Using (\ref{WarpSolution}), one
obtains
\begin{equation}
  S = -\frac{V_2}{8\pi(4\pi\kappa^2)^{2/3}}
       \int d^4x\sqrt{g^{(4)}}
       g^{(4)\mu\rho}g^{(4)\nu\sigma} F^a_{\mu\nu} F^a_{\rho\sigma} \; ,
\end{equation}
where again we denote $V_2=V(d)$.
The fact that the warp-factor contributions arising from the
4-dimensional part of the full metric cancel each other is special to
four dimensions and leads to the simple 4-dimensional hidden gauge-coupling
\begin{equation}
  \alpha_2 = \frac{(4\pi\kappa^2)^{2/3}}{2V_2} \; .
\end{equation}
Notice that the formula is the same as for the dimensional reduction
(\ref{NewtonConstant}). But here, we have to take the full quadratic
Calabi-Yau volume at hidden boundary position $x^{11}=d$ instead of
the first order linear volume. Again, we see that the hidden gauge-theory
becomes strongly coupled if the second boundary is placed in the
vicinity of the zero of the Calabi-Yau volume.

\section{The Cosmological Constant vanishes at $\kappa^{2/3}$ Order}
We saw previously that for the case with fluxes $\alpha \not= 0,
\beta_a=0$ we gain a quadratic behaviour in $x^{11}$ for the
Calabi-Yau threefold volume. The value of its zero is closely linked
to the value of the 4-dimensional Newton's Constant. Hence, it is
important to ask whether such a phenomenologically favoured value for
the modulus $d$ could be stabilized by means of an effective
4-dimensional potential. One could attack this problem by means of the
effective 4- or 5-dimensional action of heterotic M-theory derived in
[\ref{OL}]. Here we are going to examine the resulting potential for
$d$ for the derived warped graviational background by integrating out
the internal dimensions of the original 11-dimensional heterotic
M-theory action. We will find, as one expects from supersymmetry
arguments, that this potential vanishes to first order in
$\kappa^{2/3}$. Thus, in agreement with the 4-dimensional
Minkowski-solution in (\ref{WarpSolution}) the effective 4-dimensional
cosmological constant is zero at this order. Unfortunately, the
analysis at higher orders cannot be performed, yet, since it would
presuppose the complete knowledge of the heterotic M-theory action at
the appropriate orders, which is still lacking.

Let us start with the bosonic action of 11-dimensional heterotic
M-theory [\ref{HorWitt2}], which under the condition that only the
$G$-flux component $G_{a\bb c\db}$ contributes reads ($i=1,2$)
\begin{alignat}{3}
  S &= S_{11}+S_{10}^{(1)}+S_{10}^{(2)} \; , \notag \\
  S_{11} &= \frac{1}{2\kappa^2}
            \int d^{11}x \sqrt{\gh}
            \left[ -\Rh(\gh_{MN})
                   -\frac{1}{4}\Gh_{a\bb c\db}\Gh^{a\bb c\db}
            \right] \; , \notag \\
  S_{10}^{(1)} &= \frac{1}{2\kappa^2}
              \int d^{10}x \sqrt{{\gh}^{(1)}}{\cal L}^{(1)}  \; , \\
  S_{10}^{(2)} &= \frac{1}{2\kappa^2}
              \int d^{10}x \sqrt{{\gh}^{(2)}}{\cal L}^{(2)}  \; , \notag
\end{alignat}
where the hats are meant to indicate that the full warped metric is
used for determination or contractions. To leading order in
$\kappa^{2/3}$ the bosonic part of the boundary actions is given by
\begin{equation}
  {\cal L}^{(i)} = -\frac{1}{2\pi}\left(\frac{\kappa}{4\pi}\right)^{2/3}
                    \text{tr} F^{(i)}_{a\bb}F^{(i)a\bb}
                   +{\cal O}(\kappa^{4/3}) \; ,
\end{equation}
where to leading order in $\kappa^{2/3}$ the hats over the
field-strengths have been omitted since contractions must now be performed
with the original ``zero-order'' metric.
\\[-6mm]

\noindent
{\it The Measure-Factors}\\[2mm] For the case with varying volume we
found in the last section the warp-factors relation
$f(x^{11})=k(x^{11})=-b(x^{11})$. This allows to express
$\sqrt{\gh}=e^{-3b/2}\sqrt{g_{CY_3}}$ in terms of the measure on
the original Calabi-Yau threefold without warp-factors. The condition
to preserve supersymmetry gave
\begin{equation*}
  \partial_{11} (e^{-3b/2}) = -\frac{\sqrt{2}}{16}\alpha  \; ,
\end{equation*}
which together with $\alpha=8\sqrt{2}\Theta(x^{11}) {\cal S}_1$ (note that for
the case under consideration, we have $\partial_a \alpha=0$ which means
${\cal S}_1$ is a constant) leads to
\begin{equation*}
   e^{-3b(x^{11})/2} = 1-{\cal S}_1 x^{11} \; ,
\end{equation*}
thus determining the measure-factor as
\begin{equation}
  \sqrt{\gh} = (1-{\cal S}_1 x^{11})\sqrt{g_{CY_3}} \; .
\end{equation}
Analogously the boundary measures are given by $\sqrt{{\gh}^{(1)}}
=e^{-b(x^{11}=0,d)}\sqrt{g_{CY_3}}$ which leads to
\begin{equation}
  \sqrt{{\gh}^{(1)}} = \sqrt{g_{CY_3}} \; , \qquad
  \sqrt{{\gh}^{(2)}} = (1-{\cal S}_1 d)^{2/3} \sqrt{g_{CY_3}} \; .
\end{equation}
\\[-6mm]

\noindent
{\it The Curvature-Scalar}\\[2mm]
Next let us express the 11-dimensional curvature-scalar for the warp-metric
\begin{equation*}
  \Rh(\gh_{MN}) = \gh^{KL}\partial^M\partial_M \gh_{KL}
                 -\partial^K\partial^L\gh_{KL}
                 +\Gamh^P_{KL}\Gamh^Q_{MN} \gh_{PQ}
                  \left( \gh^{KL}\gh^{MN}-\gh^{KM}\gh^{LN} \right)
\end{equation*}
through the warp-factors $b,f$ and the original Calabi-Yau curvature
scalar. This gives
\begin{alignat*}{3}
         \Rh = e^{-k}
               &\bigg[ {\cal D}\, \partial_{11}^2 b 
                      +{\cal N}\, \partial_{11}^2 f 
                      +\frac{{\cal D}({\cal D}+1)}{4}(\partial_{11}b)^2
                      +\frac{{\cal N}({\cal N}+1)}{4}(\partial_{11}f)^2  \\
                     &+\frac{{\cal D}{\cal N}}{2}\partial_{11}b\,\partial_{11}f
                      -\frac{{\cal D}}{2}\partial_{11}b\,\partial_{11}k
                      -\frac{{\cal N}}{2}\partial_{11}f\,\partial_{11}k
                \bigg]
                + e^{-f}R(g_{mn}) \; ,
\end{alignat*}
where ${\cal D}$ represents the real dimension of the non-compact
external spacetime, while ${\cal N}=2n$ denotes the real dimension of
the complex $n$-dimensional internal Calabi-Yau manifold. For our
concrete case with ${\cal D}=4,\, n=3$ and $f=k=-b$ plus a Ricci-flat
Calabi-Yau manifold, we arrive at
\begin{equation*}
          \Rh = e^b 
                \left[ -2\partial_{11}^2 b
                       +\frac{5}{2}(\partial_{11}b)^2
                \right] \; .
\end{equation*}
Using $e^{-3b/2} = 1-{\cal S}_1 x^{11}$ plus the $\mathbb{Z}_2$
symmetry of the orbifold (which introduces a jump in the first
derivative of the metric at the orbifold fixed-points), we finally obtain
\begin{equation}
          \Rh = -\frac{2}{9}
                 \frac{{\cal S}_1^2}{(1-{\cal S}_1 x^{11})^{8/3}}
                +\frac{8}{3}{\cal S}_1
                      \left( -\delta(x^{11})
                             +\frac{\delta(x^{11}-d)}{(1-{\cal S}_1 d)^{5/3}}
                      \right) \; .
\end{equation}
for the curvature.
\\[-5mm]

\noindent
{\it The Field-Strengths}\\[2mm]
Let us deal next with the field-strengths appearing in the bulk and boundary
action. Using Einstein's equation
\begin{alignat}{3}
   \Rh_{MN}-\frac{1}{2}\Rh \gh_{MN} 
 = &-\frac{1}{24}\left( 4 \Gh_{MKLP}\Gh_N^{\phantom{N}KLP}
                       -\frac{1}{2}\gh_{MN}\Gh_{KLPQ}\Gh^{KLPQ}
                 \right) \\ \notag
   &-\left( \delta(x^{11}) T^{(1)}_{AB}
          +\delta(x^{11}-d) T^{(2)}_{AB}
     \right) \delta^A_M \delta^B_N
\end{alignat}
with the boundary energy-momentum tensor
\begin{alignat}{3}
 T^{(i)}_{AB} &= \frac{1}{\sqrt{g_{11,11}}}
                 \left( \frac{1}{2}\gh_{AB}{\cal L}^{(i)}
                       -\frac{\partial{\cal L}^{(i)}}{\partial\gh^{AB}}
                 \right)           \\ \notag
              &= \frac{1}{\sqrt{g_{11,11}}\,2\pi}
                 \left(\frac{\kappa}{4\pi}\right)^{2/3}
                 \left( \text{tr}F^{(i)}_{AC} F^{(i)C}_B
                       -\frac{1}{4}\gh_{AB}\text{tr}
                        F^{(i)}_{CD}F^{(i)CD}  
                 \right) 
                +{\cal O}(\kappa^{4/3}) \; ,
\end{alignat}
we may substitute the $G$-flux kinetic term in the bulk action by the
expression
\begin{equation}
    -\frac{1}{4}\Gh_{a\bb c\db}\Gh^{a\bb c\db}
  = 3\Rh-\frac{2}{3}
    \left(  \delta(x^{11})T_{\,A}^{(1)A}
           +\delta(x^{11}-d)T_{\,A}^{(2)A}
    \right)  \; .
\end{equation}
Thus the tree-level action becomes
\begin{equation}
  S = \frac{1}{2\kappa^2}
      \left\{
          \int d^{11}x \sqrt{\gh}\,2\Rh
         +\frac{1}{3}\sum_{i=1,2}\int d^{10}x\sqrt{{\gh}^{(i)}}
          \left( 2\gh^{(i)AB}\frac{\partial{\cal L}^{(i)}}
                                  {\partial\gh^{(i)AB}}
                -7 {\cal L}^{(i)}
          \right)
      \right\} \; .
\end{equation}
For the boundary contributions let us concentrate on terms with two
metric contractions which take expectation values on the Calabi-Yau
manifold. It is this class of terms which appear at $\kappa^{2/3}$
order to which we ultimately have to truncate our results. This
amounts to an additional warp-factor $e^{-2f(x^{11}=0,d)} =1/(1-{\cal
S}_1 x^{11})^{4/3}|_{x^{11}=0,d}$ from the metric contractions.

\noindent
{\it The Effective Potential}
\\[2mm] 
Putting everything together, we can now integrate over the internal
dimensions, which can be done explicitly for the eleventh
dimension\footnote{We work in the downstairs picture and employ
$\int_{-d}^d dx^{11} = 2\int_0^d dx^{11}$.}. We thus obtain
for the effective tree-level action
\begin{alignat}{3}
  S = \frac{1}{6\kappa^2}
      \int d^{10}x \sqrt{g_{CY_3}}
      \bigg\{
             &\sum_{i=1,2}\frac{1}{(1-{\cal S}_1 d^{(i)})^{2/3}}
            \left(
             2g^{lm}\frac{\partial {\cal L}^{(i)}}
                            {\partial g^{lm}}
            -7{\cal L}^{(2)}
            \right)                    \notag \\
             &+28{\cal S}_1\left( -1+\frac{1}{(1-{\cal S}_1 d)^{2/3}}
                         \right)
      \bigg\} \; ,
\end{alignat}
where we have defined $d^{(1)}=0, d^{(2)}=d$.
At this point, since the complete higher order in $\kappa^{2/3}$ terms
of the boundary actions ${\cal L}^{(i)}$ are still unknown, we have to
truncate the action to the first non-trivial $\kappa^{2/3}$ order to
proceed further. First for the boundary contributions, this truncation
gives
\begin{equation}
  2g^{lm}\frac{\partial {\cal L}^{(i)}}{\partial g^{lm}}
            -7{\cal L}^{(i)}
 = \frac{3}{2\pi}\left(\frac{\kappa}{4\pi}\right)^{2/3}
   \text{tr} F_{a\bb}^{(i)}F^{(i)a\bb} \; .
\end{equation}

To proceed further we are going to derive a relationship between both
$\text{tr}F^{(i)2}$ terms under the integral over the Calabi-Yau manifold.
Let us start with the CY-integral
\begin{equation}
  \int d^6x \sqrt{g_{CY_3}} 
  \left[ \sum_{i=1,2} \text{tr} F^{(i)}_{lm} F^{(i)lm}
        -\text{tr} R_{lm} R^{lm}
  \right] \; ,
  \label{Integral}
\end{equation}
and promote it to an integral over the $CY_3\times S^1/\mathbb{Z}_2$
manifold (we are working now at $\kappa^{2/3}$ order, which means that
the integral measure as well as the contractions and the Riemann
curvature tensor have to be taken in the original unwarped geometry)
\begin{equation}
  \int d^7x \sqrt{g_{CY_3}} 
  \left[ \sum_{i=1,2} \left(\text{tr} F^{(i)}_{lm} F^{(i)lm}
                     -\frac{1}{2}\text{tr} R_{lm} R^{lm}
                      \right) \delta(x^{11}-d^{(i)})                 
  \right]  \; .
\end{equation}
Exploiting the supersymmetry-condition resulting from the vanishing of
the gaugino-variation, $\omega^{lm} F_{lm}=0$, plus the Calabi-Yau
SU(3) holonomy condition, $\omega^{lm} R_{lm}=0$, we see that the
integral is proportional to
\begin{alignat*}{3}
  &\int d^7x \sqrt{g_{CY_3}} \,\omega^{lm}\omega^{np} 
  \bigg[  \sum_{i=1,2}
          \left( \text{tr} F^{(i)}_{[lm} F^{(i)}_{np]}
                -\frac{1}{2}\text{tr} R_{[lm} R_{np]}
          \right) \delta(x^{11}-d^{(i)})                   
  \bigg]           \\
  \propto
  &\int_{CY_3\times\mathbb{Z}_2} \, \omega \wedge
  \bigg[  \sum_{i=1,2}
          \left(\text{tr} F^{(i)} \wedge F^{(i)}
               -\frac{1}{2}\text{tr} R \wedge R
          \right) \delta(x^{11}-d^{(i)})       
  \bigg] 
  \wedge dx^{11}    \; ,
\end{alignat*}
The Bianchi-identity at this order reads [\ref{HorWitt2}]
\begin{equation*}
  dG = -\frac{1}{2\sqrt{2}\pi}
        \left( \frac{\kappa}{4\pi} \right)^{2/3}
        \sum_{i=1,2}
        \left( \text{tr} F^{(i)} \wedge F^{(i)}
              -\frac{1}{2}\text{tr}R \wedge R
        \right)
        \delta(x^{11}-d^{(i)}) \wedge dx^{11} \; , 
\end{equation*}
which says that (\ref{Integral}) is simply proportional to
\begin{equation*}
    \int_{CY_3\times\mathbb{Z}_2} \,\omega\wedge dG
   = \int_{CY_3\times\mathbb{Z}_2}  \, d(\omega\wedge G)
   = \int_{CY_3} \,\left[ \omega\wedge G \right]_{x^{11}=0}
   -\int_{CY_3} \,\left[ \omega\wedge G \right]_{x^{11}=d}
\end{equation*}
By the inverse operation as before we convert the last integral to
\begin{alignat}{3}
         \int_{CY_3} \omega\wedge G 
 \,\propto\, \int_{CY_3} d^6x \sqrt{g_{CY_3}} \,
                              \omega^{lm}\omega^{np} G_{lmnp}
 \,\propto\, \int_{CY_3} d^6x \sqrt{g_{CY_3}} \, \alpha \; .
\end{alignat}
However, since this integral in general (hence in particular for our
choice of fluxes) does not depend on $x^{11}$ at $\kappa^{2/3}$ order
[\ref{W}], the integral is the same for both boundaries thus rendering
(\ref{Integral}) vanishing.

If we neglect the $\text{tr} R^2$ contributions in (\ref{Integral})
for the low-energy boundary action because they are of higher order in
derivatives, we then obtain the desired relation
\begin{equation}
    \int_{CY_3} d^6x \sqrt{g_{CY_3}}\,\text{tr}F^{(1)2}
 = -\int_{CY_3} d^6x \sqrt{g_{CY_3}}\,\text{tr}F^{(2)2} \; .
\end{equation}
Together with the $\kappa^{2/3}$ order value for the source (again we
suppress the higher order in derivative Riemann-curvature term for the
low-energy action)
\begin{equation}
  {\cal S}_1 = \frac{1}{4\pi}\left( \frac{\kappa}{4\pi} \right)^{2/3}
               \text{tr}F_{a\bb}^{(1)}F^{(1)a\bb}
\end{equation}
this gives the following tree-level effective action
\begin{alignat}{3}
  S = \left(-1+\frac{1}{(1-{\cal S}_1 d)^{2/3}}\right)
      \frac{11}{12\pi\kappa^2}
      \left( \frac{\kappa}{4\pi} \right)^{2/3}
      \int d^{10}x \sqrt{g_{CY_3}}
      \,\text{tr} F^{(1)}_{a\bb} F^{(1)a\bb} \; .
\end{alignat}
Recognizing that the whole expression has to be truncated to
$\kappa^{2/3}$ order, the expansion of the prefactor for ${\cal S}_1 d
\ll 1$ simply results in $-1+1=0$, which shows the desired result,
namely that at this order the (tree level) cosmological constant
vanishes.

\section{The Case with General Flux: Beyond Warped Geometries}
If one wants to relax the constraint that either $\alpha$ or $\beta_a$ is zero
(which we adopted up to now), then one has to generalize the previous pure
warped geometry to a geometry which exhibits a deviation from the Calabi-Yau
metric and is not describable by a warp-factor alone. With such a generalized
Ansatz
\begin{alignat}{3}
   ds^2&= \gh_{MN} dx^M dx^N  \\
   \label{GeneralMetric}       
       &= e^{b(x^W\!\!\!,\,x^{11})}\eta_{\mu\nu}dx^\mu dx^\nu
         +\left[g_{lm}(x^n)+h_{lm}(x^n\!\!\!,\,x^{11})\right]dx^l dx^m
         +e^{k(x^W\!\!\!,\,x^{11})}dx^{11}dx^{11} \notag \; ,
\end{alignat}
we will see that the inconsistency which arose for the warp-factor $f$ if both 
$\alpha$ and $\beta_a$ were present, disappears and instead leads to
constraints on the internal spin-connection.

The $CY_3$ metric split entails a corresponding split for the internal Vielbein
$\eh^{\lb}_{\phantom{l}m} =   e^{\lb}_{\phantom{l}m}(x^n)
+ f^{\lb}_{\phantom{l}m}(x^n\!\!\!,\,x^{11})$.
Again, we will express the spin-connection through the one belonging to the
initial metric
\begin{alignat}{3}
  \Omega_{\mu\nub\lb}(\eh) &= \frac{1}{2}
           {\eh}_{\lb}^{\phantom{l}m}{\eh}_{\nub\mu}\partial_m b \; , \qquad
  \Omega_{\mu\nub\elfb}(\eh) = \frac{1}{2}\eh_{\nub\mu}
                 \eh_{\elfb}^{\phantom{11}11}\partial_{11}b \notag \\
  \Omega_{l\mb\nb}(\eh) &=  \Omega_{l\mb\nb}(e)
                          +\Omega^{(d)}_{l\mb\nb}(e,f)           \; , \qquad
  \Omega_{l\mb\elfb}(\eh) = \frac{1}{2}\eh_{\elfb}^{\phantom{11}11}
                            (\partial_{11}f_{\mb l}
                          +\eh_{\mb}^{\phantom{m}m}\eh^{\lb}_{\phantom{l}l}
                           \partial_{11}f_{\lb m})                 \\
  \Omega_{11\lb\mb}(\eh) &= \eh_{[\lb}^{\phantom{[l}l}\partial_{|11|}
                            f_{\mb]l}                            \; , \qquad 
  \Omega_{11\lb\elfb}(\eh) = -\frac{1}{2}\eh_{\lb}^{\phantom{l}l}
                 \eh_{\elfb,11}\partial_l k \notag \; ,
\end{alignat}
with all remaining terms vanishing. 
Now the deviation from the initial CY-geometry is characterised by
$f^{\lb}_{\phantom{l}m}$ and $\Omega^{(d)}_{l\mb\nb}(e,f)$. Both go to zero if
we turn off the $G$-fluxes.
Employing again the covariant constancy $D_I\eta=(\partial_I+\frac{1}{4}
\Omega_{I\Jb\Kb}(e)\Gamma^{\Jb\Kb})\eta=0$ of the original
spinor-parameter, leads to
\begin{alignat}{3}
   dx^I{\hat D}_I\etat = 
             \bigg(&- dx^u \partial_u\psi 
                    + \frac{1}{4}dx^\mu
                      \bigg[ \Gamh_\mu^{\phantom{\mu}l}\partial_l b
                            +\Gamh_\mu^{\phantom{\mu}11}\partial_{11}b
                      \bigg]
                    + \frac{1}{4}dx^l
                      \bigg[ \Omega_{lmn}^{(d)}\Gamh^{mn}
         \notag \\
                           &+2\Omega_{lm11}(\eh)\Gamh^m\Gamh^{11}
                      \bigg]
                    + \frac{1}{4}dx^{11}
                      \bigg[ \Gamh_{11}^{\phantom{11}l}\partial_l k
                            +\Omega_{11lm}(\eh)\Gamh^{lm}
                      \bigg]
             \bigg)\etat \; .
\end{alignat}
Again, specifying that our internal space consists of a Calabi-Yau
times an interval, we employ $\Gamh^{11}\etat=e^{-k/2}\etat$ and
$\Gamh^a\etat=0$, $\Gamh_{\ab}\etat=0$ plus the Dirac-algebra
$\{\Gamh^a,\Gamh^{\bb}\}=2\gh^{a\bb}$ to obtain
\begin{alignat}{3}
   dx^I{\hat D}_I\etat = \bigg\{ 
      &\bigg[ \left(\frac{1}{4}\Omega^{(d)\phantom{\,}a}_{la}
                   -\frac{1}{4}\Omega^{(d)\phantom{\,}\ab}_{l\ab}
                   -\partial_l \psi
              \right) dx^l
             +\left(\frac{1}{4}\Omega^{\phantom{11a}a}_{11a}(\eh)
                   -\frac{1}{4}\Omega^{\phantom{11\ab}\ab}_{11\ab}(\eh)
                   -\partial_{11}\psi
              \right) dx^{11}
       \bigg] \notag \\
      +&\bigg[ \frac{1}{4}e^{-k/2}\partial_{11}b\, dx_\mu\bigg]\Gamh^\mu
      +\bigg[ \frac{1}{2}e^{-k/2}\Omega_{l\ab 11}(\eh) dx^l
             -\frac{1}{4}e^{-k/2}\partial_{\ab}k dx_{11}\bigg]\Gamh^{\ab} 
                                                                    \\
      +&\bigg[ \frac{1}{4}\partial_{\ab}b\, dx_\mu \bigg]\Gamh^{\mu\ab}
      +\bigg[ \frac{1}{4}\Omega_{l\ab\bb}^{(d)} dx^l
             +\frac{1}{4}\Omega_{11\ab\bb}(\eh) dx^{11} 
       \bigg]\Gamh^{\ab\bb}
                    \bigg\}\etat 
         \label{CovariantDerivative2} \notag  \; .
\end{alignat}
For the second part of the Killing-equation which consists of the contractions
of 
$\Gamh$-matrices with the $G$-flux, it will be convenient to use the following
abbreviations
\begin{alignat}{3}
  G &= \gh^{a\bb}\gh^{c\db}G_{a\bb c\db}   \\
  G_m &= \gh^{b\cb}G_{mb\cb 11}  \\
  G_{mn} &= \gh^{c\db}G_{mnc\db}  \; .
\end{alignat}
In order to eventually extract the real and imaginary parts of the
Killing-spinor equation, we have to know their behaviour under complex
conjugation, which is given by
\begin{equation}
  \overline{G}=G \; , \quad \overline{G_a}=-G_{\ab} \; , \quad
  \overline{G^a_{\phantom{a}b}}=-G^{\ab}_{\phantom{a}\bb} \; .
\end{equation}
Analogously, to the treatment in the previous section, we then arrive at
\begin{alignat}{3}
  d&x_I\big(\Gamh^{IJKLM}-8\gh^{IJ}\Gamh^{KLM}\big)G_{JKLM}\etat
   = \bigg\{ 3e^{-k/2} \left[-24 G_{\ab}dx^{\ab}-8G_a dx^a
                             +4G dx_{11}
                       \right] \notag \\ 
            +&12 G dx_\mu\Gamh^\mu
            + 12\left[ G dx_{\ab}-6G^{\bb}_{\phantom{b}\ab}dx_{\bb}
                      +4 G_{\ab}dx^{11}
               \right]\Gamh^{\ab} 
            +24e^{-k/2} G_{\ab}dx_\mu\Gamh^{\mu\ab} 
          \label{GContractions2} \\ 
            +&12e^{-k/2}\left[2G_{\bb}dx_{\ab}
                             -3G^{\cb}_{\phantom{c}\ab\bb 11}dx_{\cb}
                       \right]\Gamh^{\ab\bb}
     \bigg\}   \notag  \; .
\end{alignat}
With (\ref{CovariantDerivative2}) and (\ref{GContractions2}) we are then able
to decompose the complete Killing-equation (\ref{Killing}) into its external,
CY- and 11-components. Thus unbroken supersymmetry finally translates into the
following constraints on the spin-connection
\begin{alignat}{3}
      \Omega^{(d)\,b}_{ab}-\Omega^{(d)\,\bb}_{a\bb} 
  &=  \frac{2\sqrt{2}}{3}e^{-k/2}G_a  \\
      \Omega^{(d)\,a}_{\phantom{(d)\,a}\bb\cb}
  &=  0 \\
      \Omega^{(d)\,c}_{\phantom{(d)\,c}ab}
  &=  \frac{\sqrt{2}}{6}e^{-k/2}
      \left( G_d\delta^{cd}_{ab}+3G^c_{\phantom{c}ab11}
      \right) \\         
      \Omega_{11a}^{\phantom{11a}a}(\eh) 
  &=  \Omega_{11\ab}^{\phantom{11\ab}\ab}(\eh) \\
      \Omega^b_{\phantom{b}\ab 11}(\eh)
  &=  \Omega_{11ab}(\eh)  
     = 0 \\
      \Omega^b_{\phantom{a}a11}(\eh)
  &=  \frac{\sqrt{2}}{12} e^{k/2} 
      \left( G\delta^b_a-6G^b_{\phantom{b}a}
      \right)
      \; ,
\end{alignat}
where
$\delta^{cd}_{ab}=\delta^c_a\delta^d_b-\delta^c_b\delta^d_a$. Additionaly, the
solution to the Killing-spinor equation provides us with
further equations, which determine the warp-factors and covariant-spinor
deviation in terms of the $G$-flux parameters
\begin{alignat}{3}
  \partial_a b &= \frac{\sqrt{2}}{3}e^{-k/2}G_a \\
  \partial_{11}b &= -\frac{\sqrt{2}}{6}e^{k/2}G 
        \label{bElevenDerivative} \\
  \partial_a k &= -\frac{2\sqrt{2}}{3}e^{-k/2}G_a \\
  \partial_a\psi &= -\frac{\sqrt{2}}{12}e^{-k/2}G_a \\
  \partial_{11}\psi &= \frac{\sqrt{2}}{24}e^{k/2}G \; .
\end{alignat}
Similarly to the last section we obtain
\begin{equation}
  8\psi(x^m,x^{11})=k(x^m,x^{11})=-2b(x^m,x^{11}) \; ,
\end{equation}
but this time a dependence on both $x^m$ and $x^{11}$ is allowed. Note that
this relation is in accordance with the result of the first order
approximation derived in [\ref{W}].

The relation $\Gamma^u_{\phantom{u}uv}(\gh)=\frac{1}{\sqrt{\gh}}\partial_v
\sqrt{\gh}$ between the Christoffel-symbols and the metric determinant enables
us to to find 
\begin{equation}
  \partial_{11}\sqrt{\gh_{CY_3}}=\sqrt{\gh_{CY_3}}
  \left( \Gamma^u_{\phantom{u}u 11}(\gh_{CY_3\times I})
        -\frac{1}{2}\partial_{11}k
  \right) \; . 
\end{equation}
Via the relation between the Christoffel-symbols and the spin-connection,
$\eh^{\ub}_{\phantom{u}x}\Gamma^x_{\phantom{x}vw}=\partial_v
\eh^{\ub}_{\phantom{u}w}+\Omega_{v\phantom{u}\xb}^{\phantom{v}\ub}
\eh^{\xb}_{\phantom{x}w}$, we get
${\Gamma^u}_{u11}(\gh_{CY_3\times I})-\frac{1}{2}\partial_{11}k
=\Omega^u_{\phantom{u}u11}(\eh_{CY_3\times I})$ and thereby
\begin{equation}
  \partial_{11}\sqrt{\gh_{CY_3}}=\sqrt{\gh_{CY_3}}
  \Omega^l_{\phantom{l}l11}(\eh) \; ,
\end{equation}
where we have used that $\Omega^{11}_{\phantom{11}1111}(\eh)=0$.
Together with the constraint on $\Omega^b_{\phantom{b}a11}(\eh)$, which gives 
$\Omega^a_{\phantom{a}a11}(\eh)=\Omega^{\ab}_{\phantom{a}\ab11}(\eh)=
\sqrt{2}\left(\frac{6+n}{12}\right)e^{k/2}G$
($n=\text{dim}_{\mathbb{C}} CY_n$), we
obtain ultimately
\begin{equation}
  \partial_{11}\ln\sqrt{\gh_{CY_3}}=\sqrt{2}\left(\frac{6+n}{6}\right)
  e^{k/2}G \; .
  \label{DerHelp}
\end{equation}
Employing the equation for $\partial_{11}b$, we can integrate this equation to
obtain the following expression for the Calabi-Yau dependence on $x^{11}$
(with $n=3$)
\begin{equation}
 V(x^{11}) = \int_{CY_3} d^6x \sqrt{\gh_{CY_3}}
           = \int_{CY_3} d^6x e^{-9b(x^{11},x^m)}C(x^m) \; ,
\end{equation}
where $C(x^m)$ arose as an integration constant by integrating (\ref{DerHelp})
over $x^{11}$. We see that now the specification of the sources simply by
means of their location in the eleventh direction is not enough to determine
$V(x^{11})$. This stems from the fact that $G$, which determines the
warp-factor $b$ contains contractions with $\gh^{a\bb}$ which itself is
$x^{11}$ dependent. Therefore the specification $G_{a\bb c\db}\propto
\Theta(x^{11}-z_i)$ does not fully determine the $x^{11}$ behaviour of
$G$. However, it is definitely true that a non-trivial $V(x^{11})$ requires
$G\not= 0$ and therefore in view of (\ref{bElevenDerivative})
$G_{a\bb c\db}\not= 0$. 

Another interesting aspect of turning on both  $G_{a\bb c\db}$ and
$G_{ab\cb 11}$ derives from the fact, that we saw above how the fluxes
determine the internal spin-connection $\Omega(\eh)$. Now it is well-known
that the spin-connection determines the holonomy-group ${\cal H}$ of a
manifold through the path-ordered exponential of $\Omega(\eh)$ around a closed
curve $\gamma$
\begin{equation}
  {\cal P} e^{\int_\gamma \Omega_m(\eh)dx^m} \in {\cal H} \; .
\end{equation} 
This is an interesting further link between the physics of $G$-fluxes and the
geometry of the compactification space. A complex 3-dimensional
K\"ahler-manifold exhibits U(3) holonomy. But we already saw above that
turning on a $G$-flux in general ruins the closedness property of the
K\"ahler-form -- therefore the new deformed manifolds are no longer
K\"ahler. This means that we do expect a more general holonomy than U(3).

\bigskip
\noindent {\large \bf Acknowledgements}\\[2ex] 
We would like to thank K.~Behrndt, S.~Gukov, D.~L\"ust, A.~Lukas,
B.A.~Ovrut and in particular E.~Kiritsis for discussion. A.K. is
supported by RTN project HPRN-CT-2000-00122.

\newpage
\begin{appendix}
\section{Appendix}
\subsection{M-Theory on a 7-Manifold}
\label{SevenMani}
In this appendix let us consider as an aside the compactification of M-theory
on a smooth 7-manifold. Our aim is to investigate which kind of internal
$G$-flux can be turned on if at the same time supersymmetry shall be preserved
and which sort of warp-factors will appear. Let us therefore start with
the warp-Ansatz
\begin{equation}
  ds^2 = \gh_{MN}dx^M dx^N
       =  e^{b(x^w)}\eta_{\mu\nu}dx^\mu dx^\nu
         +e^{f(x^w)}g_{uv}dx^u dx^v \; ,
\end{equation}
for an initial 7-manifold with metric $g_{uv}$. We will assume only an
internal nonzero G-flux. To solve the equation of motion for $G$, we write
$G=^\star \!dU$, with $U$ a 2-form and the Hodge-operation is taken with
respect to the 7-manifold. In components this reads
\begin{equation}
  G_{uvwx} = \frac{1}{4!}e^{-f/2}\epsh_{uvwxz_1z_2z_3}
              \partial^{z_1} U^{z_2 z_3} \; ,
\end{equation}
where $\epsh_{uvwxz_1z_2z_3}$ is defined as a tensor (rather than a
tensor-density) containing a factor $\sqrt{\det \gh_{uv}}$. The effect of a
nonzero $G$ on the fermionic supersymmetry-variation parameter $\eta$ will be
considered bu using $\etat = e^{-\psi(x^u)}\eta$ instead of the initial
$\eta$. Plugging this into
the Killing-spinor equation for the gravitino eventually leads to
\begin{alignat*}{3}
  \bigg( &-dx^u\partial_u\psi + \frac{1}{4}dx^\mu\Gamh_{\mu u}\partial^u b
          + \frac{1}{4}dx^u\Gamh_{u v}\partial^v f
          - \frac{\sqrt{2}}{48\times 7^3}e^{-f/2}dx^\mu\Gamh_\mu\Gamh^{uvw}
            \partial_u U_{vw} \\
         &- \frac{\sqrt{2}}{48\times 7^2}e^{-f/2}dx^u\Gamh^{vw}
            \partial_{[u} U_{vw]}
          + \frac{\sqrt{2}}{6\times 7^4}e^{-f/2}dx^u\Gamh_u^{\phantom{u}vwx}
            \partial_v U_{wx}
   \bigg) \etat  \\
 = \quad &0
\end{alignat*}
This equation is solved by
setting the various coefficients in front of the independent Gamma-matrices
to zero, which shows that the warp-factors $b,f$ and the spinor-correction
factor
$\psi$ all have to be trivially constant. In addition $\partial_{[u} U_{vw]}$
has to be zero, which amounts to a zero internal $G$-flux. Thus M-theory
compactifications on smooth 7-manifolds which are supposed to preserve some
supersymmetry do not allow for non-zero internal $G$-flux and require at the
same time trivial warp-factors.

\subsection{$G$-Flux Contractions}
\label{FluxAppendix}
First, we present those identities, which are used in the evaluation of the
Killing-spinor equation
\begin{alignat}{3}
  \Gamh^{ea\bb c\db}\etat &= 0 \notag \\
  \Gamh^{\eb a\bb c\db}\etat &= \big[ 
                                  \Gamh^{\eb}(\gh^{a\bb}\gh^{c\db}
                                             -\gh^{a\db}\gh^{c\bb})
                                 +\Gamh^{\bb}(\gh^{a\db}\gh^{c\eb}
                                             -\gh^{a\eb}\gh^{c\db})
                                 +\Gamh^{\db}(\gh^{a\eb}\gh^{c\bb}
                                             -\gh^{a\bb}\gh^{c\eb}) 
                                \big] \etat \notag \\
  \Gamh^{abc\db}\etat &= 0  \qquad\qquad\qquad\qquad  \notag \\
  \Gamh^{a\bb c\db}\etat &= \big[
                            \gh^{a\bb}\gh^{c\db}-\gh^{a\db}\gh^{c\bb}
                            \big]\etat 
                                            \notag \\
  \Gamh^{\ab\bb c\db}\etat &= \big[ 
                               \Gamh^{\ab}\Gamh^{\bb}\gh^{c\db}
                               -\Gamh^{\ab}\Gamh^{\db}\gh^{c\bb}
                               +\Gamh^{\bb}\Gamh^{\db}\gh^{c\ab} 
                              \big]\etat 
                                            \notag \\
  \Gamh^{ab\cb 11}\etat &= 0 \qquad\qquad\qquad\qquad \notag \\
  \Gamh^{\ab b \cb 11}\etat &= e^{-h/2}\big[ 
                                         \Gamh^{\ab}\gh^{b\cb}
                                        -\Gamh^{\cb}\gh^{b\ab} 
                                       \big]\etat 
                                            \notag \\
  \Gamh^{ab\cb}\etat &= 0 \qquad\qquad\qquad \notag \\
  \Gamh^{a\bb\cb}\etat &= \big[
                          \gh^{a\bb}\Gamh^{\cb}-\gh^{a\cb}\Gamh^{\bb}
                          \big]\etat
                                            \notag \\
  \Gamh^{a\bb}\etat &= \gh^{a\bb}\etat       \; .
                 \label{GammaIdentities}
\end{alignat}
With their help and the definitions (\ref{Alpha}),(\ref{Beta}),(\ref{GData}),
we arrive at the contractions 
\begin{alignat}{3}
  \Gamh^{\mu uvwx}G_{uvwx}\etat &= 
        -3\big[ e^{-2f}\alpha\Gamh^\mu
               +4ie^{-k/2-f}\beta_{\ab}\Gamh^\mu
               \Gamh^{\ab}\big]\etat \notag \\
  \Gamh^{e uvwx}G_{uvwx}\etat &=
        -12ie^{-k/2-f}\beta^e\etat \notag \\
  \Gamh^{\eb uvwx}G_{uvwx}\etat &=
        -3\big[ e^{-2f}\alpha\Gamh^{\eb}
              -4ie^{-f}\Theta^{\eb}_{\phantom{e}\ab}\Gamh^{\ab}
              -4ie^{-k/2-f}\beta^{\eb} \notag \\
               &\phantom{=\qquad}
              +4ie^{-k/2-f}\beta_{\ab}\Gamh^{\eb}\Gamh^{\ab}
              +4e^{-k/2}\Gamh^{\ab}\Gamh^{\bb}G^{\eb}_{\phantom{e}\ab\bb 11}
         \big]\etat \notag \\
  \Gamh^{11 uvwx}G_{uvwx}\etat &= -3e^{-k/2-2f}\alpha\etat
       \label{FiveGammaContractions}
\end{alignat}
and
\begin{alignat}{3}
  \gh^{\mu u}\Gamh^{vwx}G_{uvwx}\etat &= 0 \notag \\
  \gh^{au}\Gamh^{vwx}G_{uvwx}\etat &= -3ie^{-k/2-f}\beta^a\etat \notag \\
  \gh^{\ab u}\Gamh^{vwx}G_{uvwx}\etat &= -3\big[ie^{-k/2-f}\beta^{\ab}
                    +ie^{-f}\Theta^{\ab}_{\phantom{a}\eb}\Gamh^{\eb}
                    -e^{-k/2}\Gamh^{\bb\cb}G^{\ab}_{\phantom{a}\bb\cb 11}
                                          \big]\etat \notag \\
  \gh^{11 u}\Gamh^{vwx}G_{uvwx}\etat &= 3ie^{-f}\beta_{\cb}\Gamh^{\cb}
                    \gh^{11,11}\etat
       \label{ThreeGammaContractions} \; ,
\end{alignat}
which are used in the main text.

\subsection{The Heterotic String with Torsion}
\label{StringCase}
The Ansatz traditionally used [\ref{CHSW}]
in compactifications of the 10-dimensional heterotic $E_8\times E_8$
string on $CY_3$ to four dimensions with ${\cal N}=1$ supersymmetry is to make
the susy-variations of the gravitino $\psi_M$, the dilatino $\lambda$ and
the gluino $\chi ^a$ ($i,j=5,\hdots,10$)
\begin{alignat}{3}
  \delta \psi_i&=\frac{1}{\kappa}D_i\eta +\frac{\kappa}{32g^2\phi}
  ({\Gamma_i}^{jkl}-9\delta_i^j\Gamma^{kl})\eta H_{jkl}\nonumber\\
  \delta \lambda&=-\frac{1}{\sqrt{2}\phi}(\Gamma \cdot \partial \phi)
  \eta + \frac{\kappa}{8\sqrt{2}g^2\phi}\Gamma^{ijk}\eta H_{ijk}\\
  \delta \chi&=-\frac{1}{4g\sqrt{\phi}}\Gamma^{ij}F_{ij}\eta \nonumber
\end{alignat}
vanish by assuming that $H=d\phi=0$. Here $\phi$ is the dilaton and 
$H$ the
gauge-invariant field strength of the NSNS 2-form $B$, which in addition has
to fulfil the Bianchi identity
\begin{equation}
dH=\text{tr} R\wedge R - \frac{1}{30}\text{tr} F\wedge F \; .
\end{equation}
This leads to the consequence that $CY_3$ is a K\"{a}hler manifold with
$c_1(CY_3)=0$ and $SU(3)$ holonomy (and the gauge field $A$ being a
holomorphic connection on a holomorphic vector bundle $V$ over the
Calabi-Yau threefold $CY_3$ obeying the Donaldson-Uhlenbeck-Yau
equation). 

This Ansatz was generalized in [\ref{S}] to include a
non-vanishing torsion $H\neq 0$ where solutions leading again to ${\cal N}=1$
supersymmetry in $D$=4 were obtained by allowing for a warp-factor $e^{2D(y)}$
in the metric (in Einstein-frame)
\begin{equation}
    g^E_{AB}(x,y) = e^{2D(y)}g_{AB}(x,y) = e^{2D(y)}
      \left( 
      \begin{array}{cc}
          \eta_{\mu\nu} &   0      \\
                 0      & g_{mn}(y)
      \end{array}
      \right) \; ,
  \label{StromingerMetric}
\end{equation}
where we denote 10-dimensional indices by $A,B,C,\hdots$. It turns out that
$D$ has to be the dilaton $\phi$. The torsion and the dilaton are determined by
\begin{alignat}{3}
  H&=&\frac{i}{2}(\bar{\partial}-\partial) J 
                           \label{StromingerTorsion} \\
  e^{8\phi}&=&e^{8\phi_0}||\Omega ||    \; ,
\end{alignat}
where the fundamental (1,1) form $J$ is built out of the complex
structure ${J_m}^n$ as $J=\frac{1}{2}{J_m}^n g_{np}dy^m\wedge
dy^p=ig_{a\bar{b}} dz^a\wedge d\bar{z}^{\bar{b}}$ (in our conventions $J$
equals up to a minus-sign
the K\"{a}hler-form $\omega$) and $\Omega$ is the (determined up to an overall
constant) holomorphic 3-form with norm 
$||\Omega ||=(\Omega_{a_1a_2a_3}\bar{\Omega}_{\bar{b}_1\bar{b}_2\bar{b}_3}
g^{a_1\bar{b}_1}g^{a_2\bar{b}_2}g^{a_3\bar{b}_3})^{1/2}$. To recognize the
relation between $H$ (commonly called torsion) and the original torsion, we
note that the metric torsion of a complex manifold is
specified by the expression
\begin{equation}
  T^a_{bc} = -2g^{\db a}g_{\db[b,c]}
\end{equation}
and its complex conjugate. Hence, the above expression for $H$ can be
explicitely 
expressed through the metric torsion via
\begin{equation}
  H=\frac{1}{4}\left( T_{a\bb\cb}dz^a\wedge dz^{\bb}\wedge dz^{\cb}
                     +T_{\cb ab}dz^a\wedge dz^b\wedge dz^{\cb} \right)
   \; .
\end{equation}
Finally, the link between $H$ and the warp-factor is given implicitly by the
dilatino equation
$\delta\lambda =0$, which manifests itself in the following relationship
\begin{alignat}{3}
   d^\dagger J&=i(\partial-{\bar \partial}) \ln ||\Omega||\nonumber\\
              &=8i(\partial-{\bar \partial})(\phi - \phi_0) \; .
\label{nonKaehler}
\end{alignat}
From the left-hand side of this equation it can be easily discerned, that the
right-hand side serves as a measure for the non-K\"{a}hlerness of the 
compactification manifold. Therefore, by turning on $H$-torsion, the
compactification manifold becomes deformed to a manifold which is no longer
K\"ahler.

To gain a more explicit relation between the $H$-torsion and the resulting
warp-factor, we note that the dilatino equation $\delta\lambda = 0$ can be
alternatively written as [\ref{S}]
\begin{equation}
 8\partial_m \phi = {J_m}^n \nabla_p {J_n}^p \; .
\end{equation} 
Here, the covariant derivative is constructed out of the initial metric
$g_{MN}$ without warp-factor. The $H$-covariant constancy of the complex
structure [\ref{S}]
\begin{equation}
  \nabla_m {J_n}^p - {H_{qm}}^p {J_n}^q - {H^q}_{mn} {J_q}^p = 0
\end{equation}
plus its property to square to minus the identity,
${J_m}^n{J_n}^p=-\delta^p_m$, serve together with $J_{a\bb}=ig_{a\bb}$ to
derive
\begin{equation}
  8\partial_a \phi = {H_{ab}}^b-{H_{a\bb}}^{\bb} \; .
\end{equation}
The contraction is with respect to the initial metric in whose frame the
relation holds. Equation (\ref{StromingerTorsion}), which relates $H$-torsion
with metric torsion, reads in components $H_{ab\cb}=-g_{\cb[a,b]}$ and leads
to ${H_{a\bb}}^{\bb}=-{H_{ab}}^b$. Finally, to obtain the relation between the
warp-factor $\phi$ and $H$-torsion in the Einstein-frame, we have to transform
the contractions according to the rescaling $g_{AB}=e^{-2\phi}g_{AB}^E$ from
the initial frame to the Einstein-frame and gain
\begin{equation}
  \partial_a(e^{-2\phi})=-\frac{1}{2}{H_{ab}}^b \; .
      \label{WarpTorsion}
\end{equation}
The contraction on the right-hand-side is now understood to be carried out
with $g^E_{AB}$. 

Furthermore warp-geometries appear in heterotic five-brane solutions
preserving supersymmetry. They were obtained 
([\ref{Shet}],[\ref{CHS1}],[\ref{CHS2}]; cf. also the axionic instantons in
[\ref{KKL}]) with the Ansatz ($k,l,m,n=7,\hdots,10$)
\begin{alignat*}{3}
g_{mn}&=e^{2\phi}\delta_{mn}  \\
H_{mnl}&=-\epsilon_{mnl}^{\phantom{mnl}k}\partial_k\phi
\end{alignat*}
showing again that turning on torsion leads to a warp factor.

\end{appendix}

\section*{References}
\begin{enumerate}

\item
\label{CHSW}
P. Candelas, G. Horowitz, A. Strominger and E. Witten, {\em Vacuum
Configurations for Superstrings}, Nucl. Phys. {\bf B258} (1985) 46.

\item
\label{Spence}
B.S.~Acharya and B.~Spence,
{\em Flux, Supersymmetry and M-Theory on Seven-Manifolds},
hep-th/0007213.

\item
\label{S}
A. Strominger, {\em Superstrings with Torsion}, Nucl. Phys. {\bf B274}
(1986) 253.

\item
\label{Shet}
A. Strominger, {\em Heterotic Solitons},  Nucl. Phys. {\bf B343}
(1990) 167.

\item
\label{CHS1}
C. Callan, J. Harvey and A. Strominger, {\em Worldsheet Approach to
Heterotic Instantons and Solitons}, Nucl. Phys. {\bf B359} (1991) 611.

\item
\label{CHS2}
C. Callan, J. Harvey and A. Strominger, {\em Worldbrane Actions for
String Solitons}, Nucl. Phys. {\bf B367} (1991) 60.

\item
\label{KKL}
C. Kounnas, E. Kiritsis and D. L\"ust, {\em A Large Class of New
Gravitational and Axionic Backgrounds for Four-dimensional Superstrings},
Int.J.Mod.Phys. A9 (1994) 1361,
hep-th/9308124. {\em Non-compact Calabi-Yau Spaces and other
Non-trivial Backgrounds for 4D-Superstrings}, "Essays on Mirror Manifolds, II",
hep-th/9312143.

\item
\label{W}
E. Witten, {\em Strong Coupling Expansion of Calabi-Yau
Compactification}, Nucl. Phys. {\bf B471} (1996) 135, hep-th/9602070.

\item
\label{BG}
K. Behrndt and S. Gukov, {\em Domain Walls and Superpotentials from M-Theory
on Calabi-Yau Threefolds}, Nucl. Phys. {\bf B580} (2000) 225, hep-th/0001082.

\item
\label{StromLoop} 
A. Strominger, {\em Loop Corrections to the Universal
Hypermultiplet}, Phys. Lett. {\bf B421} (1998) 139, hep-th/9706195.

\item
\label{WittenM}
E. Witten, {\em String Theory Dynamics in Various Dimensions},
Nucl.Phys. B443 (1995) 85; hep-th/9602070.

\item
\label{HorWitt1}
P. Ho\v{r}ava and E. Witten, {\em Heterotic and Type I String Dynamics 
from Eleven Dimensions}, Nucl.Phys. {\bf B460} (1996) 506, hep-th/9510209.

\item
\label{HorWitt2}
P. Ho\v{r}ava and E. Witten, {\em Eleven-Dimensional Supergravity on a
Manifold with Boundary},
Nucl.Phys. {\bf B475} (1996) 94; hep-th/9603142.

\item
\label{OL}
A. Lukas, B.A. Ovrut, K.S. Stelle and D. Waldram,
{\em Heterotic M-Theory in Five-Dimensions},
Nucl. Phys. {\bf B552} (1999) 246, hep-th/9806051.\\
A. Lukas, B.A. Ovrut and D. Waldram,
{\em On the Four-Dimensional Effective Action of Strongly Coupled
Heterotic String Theory},
Nucl. Phys. {\bf B532} (1998) 43, hep-th/9710208.

\item
\label{BB}
K. Becker and M. Becker, {\em M-Theory on Eight-Manifolds}, 
Nucl.Phys. {\bf B477} (1996) 155, hep-th/9605053.

\item
\label{DRS}
K. Dasgupta, G. Rajesh and S. Sethi, {\em M Theory, Orientifolds 
and G-Flux}, JHEP 9908 (1999) 023, hep-th/9908088.

\item
\label{CPV}
C. S. Chan, P. L. Paul and H. Verlinde, {\em A Note on 
Warped String Compactification}, hep-th/0003236.

\item
\label{GSS}
B. R. Greene, K. Schalm and G. Shiu, {\em Warped 
Compactifications in M and F Theory}, hep-th/0004103.

\item
\label{BB2}
K.~Becker and M.~Becker, {\em Compactifying M-Theory to Four-Dimensions},
hep-th/0010282.

\item
\label{KB}
K~Becker, {\em A Note on Compactifications on Spin(7)-Manifolds},
hep-th/0011114.

\item
\label{GSS2}
B. R. Greene, K. Schalm and G. Shiu, {\em Dynamical Topology
Change in M Theory}, hep-th/0010207.

\end{enumerate}

\end{document}